\documentclass[11pt,a4paper]{article}
\pdfoutput=1
\usepackage{jcappub}
\usepackage{rotating}
\usepackage{array}
\usepackage{amsmath}
\usepackage[normalem]{ulem}
\usepackage{slashed}
\usepackage{bm}
\usepackage{booktabs}
\usepackage[pdftex,table,dvipsnames]{xcolor}
\usepackage{units}
\usepackage[all]{hypcap}

\def \apj   {ApJ}

\renewcommand{\arraystretch}{1.2}

\usepackage{multirow}

\usepackage[pdftex]{pict2e}

\newcommand{\diff}{\mathrm{d}}

\arxivnumber{TTK-23-02, TTP-23-008, LAPTH-008/23}

\title{Fast and accurate AMS-02 antiproton likelihoods for global dark matter fits}

\author[a,b,1]{Sowmiya Balan,\note{Corresponding authors}}
\author[a,b]{Felix Kahlhoefer,}
\author[c]{Michael Korsmeier,}
\author[a,d]{Silvia Manconi}
\author[a,1]{and Kathrin Nippel}

\affiliation[a]{Institute for Theoretical Particle Physics and Cosmology (TTK), RWTH Aachen University, D-52056 Aachen, Germany}
\affiliation[b]{Institute for Theoretical Particle Physics (TTP), Karlsruhe Institute of Technology (KIT),\\ 76128
Karlsruhe, Germany}
\affiliation[c]{The Oskar Klein Centre for Cosmoparticle Physics, Department of Physics, Stockholm University, Alba Nova, 10691 Stockholm, Sweden}
\affiliation[d]{Laboratoire d'Annecy-le-Vieux de
Physique Théorique (LAPTh), CNRS, USMB, F-74940 Annecy, France}

\emailAdd{sowmiya.balan@kit.edu}
\emailAdd{kahlhoefer@kit.edu}
\emailAdd{michael.korsmeier@fysik.su.se}
\emailAdd{manconi@lapth.cnrs.fr}
\emailAdd{nippel@physik.rwth-aachen.de}

\abstract{The antiproton flux measurements from AMS-02 offer  valuable information about the nature of dark matter, but their interpretation is complicated by large uncertainties in the modeling of cosmic ray propagation. In this work we present a novel framework to efficiently marginalise over propagation uncertainties in order to obtain robust AMS-02 likelihoods for arbitrary dark matter models. The three central ingredients of this framework are: the neural emulator \textsf{DarkRayNet}, which provides highly flexible predictions of the antiproton flux; the likelihood calculator \textsf{pbarlike}, which performs the marginalisation, taking into account the effects of solar modulation and correlations in AMS-02 data; and the global fitting framework \textsf{GAMBIT}, which allows for the combination of the resulting likelihood with a wide range of dark matter observables. We illustrate our approach by providing updated constraints on the annihilation cross section of WIMP dark matter into bottom quarks and by performing a state-of-the-art global fit of the scalar singlet dark matter model, including also recent results from direct detection and the LHC.  }


\begin{document}

\maketitle
\flushbottom

\section{Introduction}

The search for dark matter (DM) is a global effort, both in the geographical sense, i.e.\ in terms of the sheer number of experiments and groups involved, and in the methodological one, i.e.\ in terms of the variety of relevant constraints that need to be considered. The scope of this endeavour calls for computational tools that provide fast and accurate theory predictions of relevant observables, efficient routines for likelihood calculations, and methods for their statistical interpretation. Of particular importance are tools that can perform DM relic density calculations and combine the results with a variety of constraints from direct and indirect detection experiments, such as \textsf{micrOmegas}~\cite{Belanger:2018ccd}, \textsf{DarkSUSY}~\cite{Bringmann:2018lay}, \textsf{MadDM}~\cite{Ambrogi:2018jqj}, \textsf{SuperIso Relic}~\cite{Arbey:2018msw} and the \textsf{DarkBit}~\cite{GAMBITDarkMatterWorkgroup:2017fax} module of the \textsf{GAMBIT}~\cite{GAMBIT:2017yxo} global fitting framework.

An excellent example for the need for tools are satellite measurements of cosmic rays (CRs), which are sensitive to charged antiparticles produced in DM annihilations \cite{Bergstrom:1999jc,Donato:2003xg,Bringmann:2006im,Donato:2008jk,Fornengo:2013xda,Evoli:2011id,Bringmann:2014lpa,Cirelli:2014lwa,Cembranos:2014wza,Hooper:2014ysa,Boudaud:2014qra,Giesen:2015ufa,Reinert:2017aga,Evoli:2015vaa,Luque:2021ddh}. In particular the most recent measurements of the antiproton flux by AMS-02~\cite{Aguilar:2021tos} have become so precise that we can expect strong constraints on, or potential evidence for,  many different DM models, provided the background from secondary antiprotons produced in astrophysical systems can be modelled with sufficient accuracy~\cite{2103.09824,Calore:2022stf}.

Unfortunately, a detailed modeling of CR propagation in the Galaxy needed for this purpose is computationally expensive. 
The state-of-the-art modeling for the CR journey in our Galaxy solves such transport problem numerically.  
Many astrophysical ingredients enter the computation, such as CR injection and transport properties. This requires a large number of parameters to be included in the analysis, some of them poorly constrained. While it is possible to consider these uncertainties in the calculation of AMS-02 antiproton constraints~\cite{Cuoco:2017iax}, including these constraints in global fits of DM models together with direct detection and collider constraints is very challenging without simplifying assumptions (see Ref.~\cite{Arina:2019tib}). 

In the present work, we address this gap in the toolbox of DM phenomenologists by providing three separate codes that together form a full analysis chain for AMS-02 antiproton data in the context of analysing DM models:
\begin{itemize}
    \item \textsf{DarkRayNet.v2} is an updated version of the tool first presented in Ref.~\cite{2107.12395}. It provides fully trained neural networks that predict the primary (i.e.\ DM-produced) and secondary antiproton fluxes as a function of the DM properties and a wide range of propagation parameters. Compared to the first release, \textsf{DarkRayNet.v2} contains more flexible propagation models and additional training data.
    \item \textsf{pbarlike} takes predictions for the antiproton flux in order to calculate the likelihood of the AMS-02 measurements, including a state-of-the-art treatment of correlations. Moreover, it provides the marginalised likelihood when the propagation parameters and the effect of solar modulation are treated as nuisance parameters.
    \item \textsf{GAMBIT 2.4} is the most recent version of the \textsf{GAMBIT}~\cite{GAMBIT:2017yxo} global fitting framework. It includes the necessary interface to calculate these likelihoods in large-scale parameter scans in order to perform global fits of many different DM models.\footnote{For additional features of the new \textsf{GAMBIT} release, we refer to Refs.~\cite{DMsimp,gravitino}.}
\end{itemize}

We illustrate the usefulness of these codes in two applications. First, we provide updated model-independent constraints on the annihilation cross section of DM into bottom quarks. In agreement with recent studies~\cite{1712.00002,1903.01472,2005.04237,Calore:2022stf}, we show that the previously observed excess around DM masses of 100 GeV~\cite{Cuoco:2016eej,Cui:2016ppb,Cholis:2019ejx} is reduced to negligible significance when correlations and uncertainties are properly included. Moreover, the nature of the excess depends on the specific propagation model being used, highlighting the need for a sufficiently flexible propagation model.

As a second application, we provide the most up-to-date global analysis of the scalar singlet DM model~\cite{Silveira:1985rk,McDonald:1993ex,Burgess:2000yq}, which has been the subject of a number of global analyses in recent years~\cite{1306.4710,1705.07931,1806.11281}. In addition to the AMS-02 antiproton data, we also include the most recent measurements of the Higgs invisible width from the LHC~\cite{2201.11585,2202.07953} and new constraints from the direct detection experiments LZ~\cite{LZ:2022ufs} and PandaX-4T~\cite{2107.13438}. We find that with these constraints, the high-mass region ($m_s > m_h$) is largely excluded and only the resonance region ($m_s \approx m_h/2$) remains of interest.

The remainder of this work is structured as follows. In section~\ref{sec:CRs} we present the framework for predicting the CR flux and introduce the specific injection and propagation models that we consider. The AMS-02 likelihood implementation and the formalism for marginalisation of nuisance parameters, and results of model-independent parameter scans are discussed in section~\ref{sec:ams_likelihood}. Finally, we introduce the model of scalar singlet DM in section~\ref{sec:ssdm} and provide our results from a global analysis of this model. Additional details on the numerical tools are provided in the appendix.

\section{Injection and propagation of cosmic rays}
\label{sec:CRs}

The transport of CRs in the Galaxy can be described by the following equation \cite{Strong:2007nh}:
\begin{eqnarray}
  \label{eq:propagation_equation}
  \frac{\partial \psi_i (\bm{x}, p, t)}{\partial t} = 
    q_i(\bm{x}, p) &+&  
    \bm{\nabla} \cdot \left(  D_{xx} \bm{\nabla} \psi_i - \bm{V} \psi_i \right)  \\ \nonumber
     &+&  \frac{\partial}{\partial p} p^2 D_{pp} \frac{\partial}{\partial p} \frac{1}{p^2} \psi_i - 
    \frac{\partial}{\partial p} \left( \frac{\diff p}{\diff t} \psi_i  
    - \frac{p}{3} (\bm{\nabla \cdot V}) \psi_i \right) -
    \frac{1}{\tau_{f,i}} \psi_i - \frac{1}{\tau_{r,i}} \psi_i \; .
\end{eqnarray}
where $\psi_i (\bm{x}, p, t)$ is the CR number density per volume and absolute momentum $p$, for each CR species $i$, evaluated at position $\bm{x}$ in Galactocentric coordinates. 
The various terms in eq.~\eqref{eq:propagation_equation} are carefully explained in the remainder of this section.

We employ \textsf{Galprop version 56}~\cite{Strong:1998fr} and \textsf{Galtoollibs 855}\footnote{ \href{https://galprop.stanford.edu}{https://galprop.stanford.edu} }  to solve numerically the transport equations for each species required in the following, such as for the training of \textsf{DarkRayNet}, see section~\ref{sec:Multinest}.
A number of custom modifications to the public \textsf{Galprop} code have been implemented as described in Ref.~\cite{1903.01472}.
The \textsf{Galprop} code permits state-of-the-art modeling of CR transport in the Galaxy and is widely used to compute CR spectra at Earth's position, as well as CR non-thermal emission at different wavelengths. 
Other, fully numerical codes such as \textsf{DRAGON} \cite{Evoli:2008dv,Evoli:2017vim}, \textsf{PICARD} \cite{Kissmann:2014sia}, as well as semi-analytical frameworks such as \textsf{USINE} \cite{Maurin:2018rmm} are also used in the recent literature.  
CRs are assumed to be in steady state (left-hand side in eq.~\eqref{eq:propagation_equation} is zero). 
The equations are solved on a 3-dimensional grid, where two dimensions describe the CR's spatial distribution, and the remaining one their kinetic energy. 
The spatial grid is defined in cylindrical coordinates, being $r$ the radial distance from the Galactic center and $z$ the distance perpendicular to the plane. The kinetic energy grid is logarithmically spaced, and the ratio between successive grid points is set to 1.3. The values of the step sizes for the spatial grid follow our previous work \cite{2107.12395}.

\subsection{Primary and secondary cosmic rays}

The term $q_i(\bm{x}, p)$ in eq.~\eqref{eq:propagation_equation} incorporates the injection of CRs at each Galactic position $\bm{x}$, with a given energy dependence.
Depending on the considered CR species $i$, we include  primary and/or secondary contributions, which are briefly illustrated in the next subsections. 
We define as primary CR sources both the standard astrophysical sources such as supernova remnants (section~\ref{subsec:primary_pHe}), along with more exotic CR injections such as the production of (anti)particle CRs in the annihilation or decay of DM particles in the Galactic DM halo (section~\ref{subsec:primary_dm}).   
Secondary CRs are instead produced by the interaction of the primary CRs with the gas in the interstellar medium (ISM) through fragmentation or decay processes 
(section~\ref{subsec:sec}). 

\subsubsection{Primary p, He}\label{subsec:primary_pHe}
CRs are dominated by protons, which account for about 90\% of the observed flux, followed by 10\% of helium (He) and even smaller percentages of heavier nuclei, electrons and positrons \cite{Aguilar:2021tos}.   
In the standard paradigm, Galactic CR  such as protons, electrons, He and other nuclei are accelerated and then released in primary sources such as supernova remnants~\cite{Gabici:2019jvz}. 
Diffusive shock acceleration is considered to be the main mechanism for promoting these particles up to multi-TeV energies. 
Particle-in-cell simulations have demonstrated that this mechanism produces power-law spectra for the accelerated particles \cite{Park:2014lqa}, which can be modified during their escape from the remnant \cite{Celli:2019qcs,Morlino:2021rzv}. 
In addition, a break in the injection spectrum at low energies is required to explain data at about few GeV, which could be connected to self-confinement of particles in supernova remnants \cite{Jacobs:2021qvh}. 

\begin{figure}[t]
\setlength{\unitlength}{0.01\textwidth}
\begin{picture}(100,56)
 \put(4,  28){\includegraphics[width=.90\textwidth] {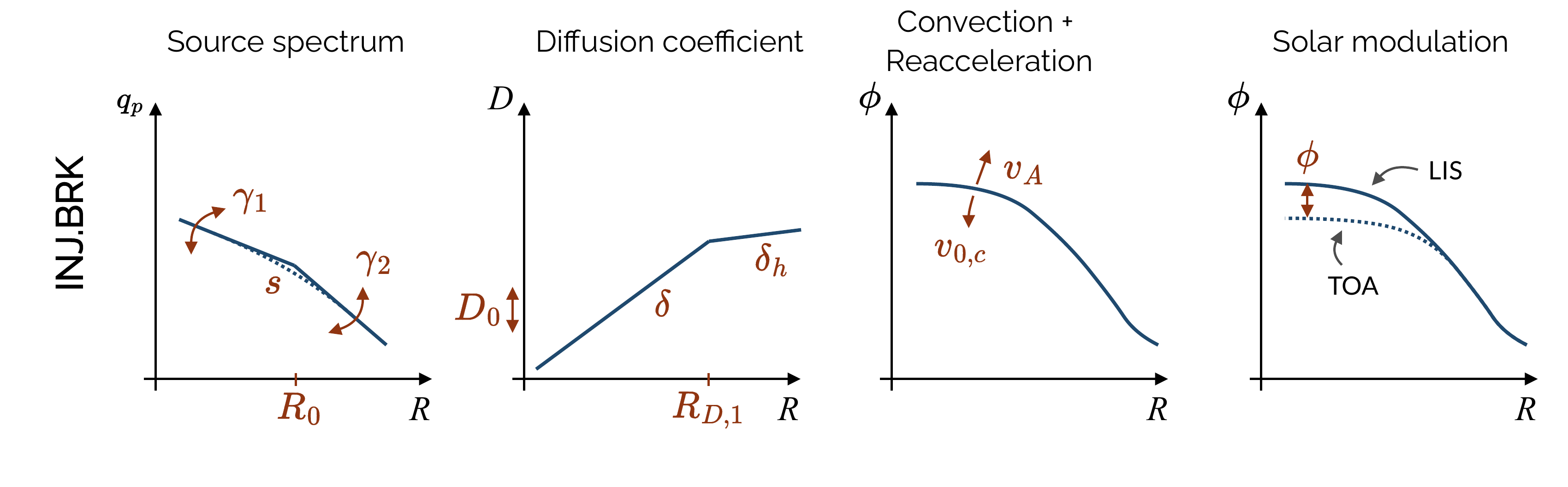}}
 \put(4,  -2){\includegraphics[width=.90\textwidth]
{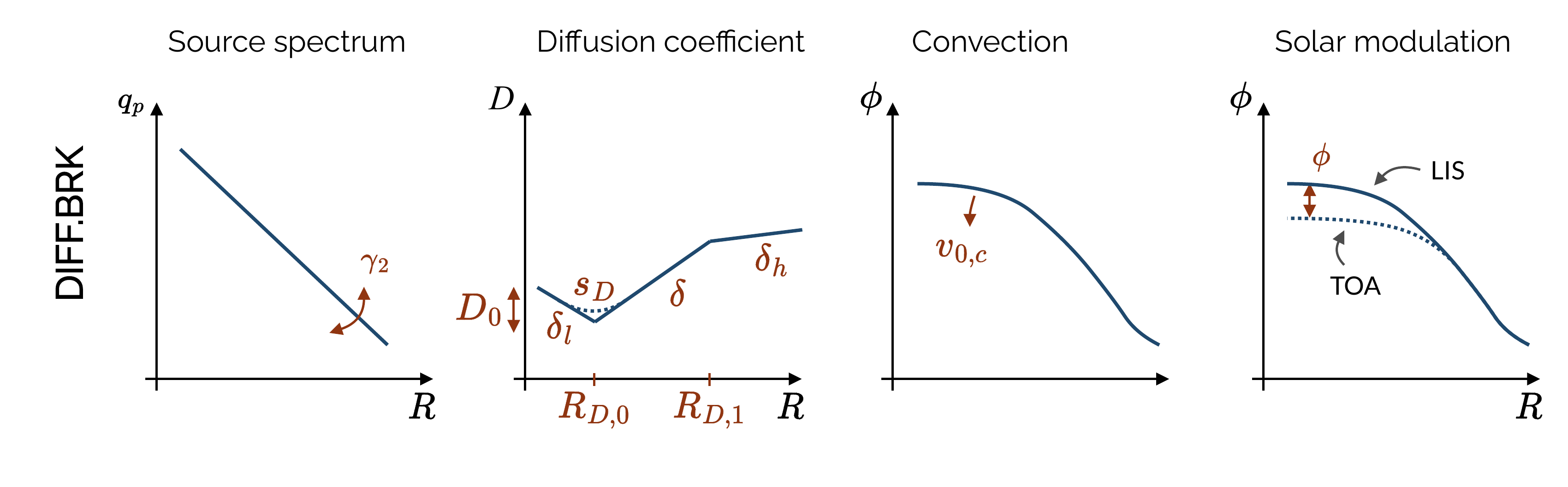}}
  \put( 3,    28){\color{gray} \line(1,0){94}}
\end{picture}
		\caption{
            Sketch of the propagation setups INJ.BRK (upper row)  and DIFF.BRK (lower row).  The different panels depict the effect of the model parameter on the spectrum $\phi$ or individual terms of the transport equation~\ref{eq:propagation_equation}.
            }
		\label{fig:prop_models}
\end{figure}

We model the source term for primary $p$ and He by factorizing the spatial and energetic dependence. 
The first is characterized by a smooth distribution of supernova remnants in Galactocentric cylindrical coordinates ($r,z$, with $r=\sqrt{x^2 +y^2}$) of the form 
$f(r, z) =\left(\frac{r}{r_{\odot}}\right)^{\alpha_1} \exp(-\alpha_2(r-r_{\odot})/r_{\odot})) \exp({-\mid z\mid/z_0})$, 
where the parameters $\alpha_1=1.09, \alpha_2=3.87, z_0=0.2$  are fixed according to default prescriptions of \textsf{Galprop}, and the Earth's distance from the Galactic Center is set to $r_{\odot}=8.5$~kpc. The modeling of the spatial distribution of sources of CR nuclei has been demonstrated to have a very minor impact on the resulting fluxes at Earth~\cite{Korsmeier:2016kha}. 
The energetic dependence is modeled as a power law or as a smoothly broken power law, depending on the chosen setup (see section~\ref{sec:models}). This approximation is widely used \cite{Boschini:2017fxq,Korsmeier:2021bkw,2012.12853,Luque:2021joz} and permits to well describe the data in the rigidity range we consider:
\begin{eqnarray}\label{eq:psp}
    \label{eq:SourceTerm_2}
    q_R(R)     &=&   \left( \frac{R}{R_0} \right)^{-\gamma_1}
                     \left( \frac{R_0^{1/s}+R^{1/s} }
                                 {2\,R_0^{1/s}      } \right)^{-s (\gamma_2-\gamma_1)},
\end{eqnarray}
which is given  as a function of rigidity $R$ (momentum $p$ divided by the absolute charge value), a  more natural quantity than momentum when dealing with charged particle acceleration. 
As illustrated by figure~\ref{fig:prop_models}, when we consider a simple power law, only one spectral index $\gamma_2$ is relevant. When a break in the injection spectrum at the rigidity value $R_0$ is considered, this is regulated by a smoothing parameter $s$, and requires two spectral indices below ($\gamma_1$) and above ($\gamma_2$) the break. 
We assume a universal injection spectrum  for all primary nuclei, except for protons for which different spectral indices $\gamma_{1,p}, \gamma_{2,p}$ are introduced, see discussion in  \cite{Korsmeier:2021bkw} and references therein.

\subsubsection{Primary antiprotons from dark matter}\label{subsec:primary_dm}
DM interactions in the diffusion halo of our Galaxy could inject further, exotic primary CRs through annihilation or decay. 
These processes produce an equal amount of matter and antimatter CRs, such as protons and antiprotons. However, in the standard paradigm the antimatter CRs such as antiprotons, antideuterons are produced only as secondaries (see next section), and their production is suppressed  with respect to primaries by 4-5 orders of magnitude. Thus antimatter CR fluxes are sensitive targets to search for exotic spectral components \cite{2010pdmo.book..521S}. 
Further astrophysical mechanism have been explored in the literature to produce primary  antimatter, such as the acceleration  in old SNRs \cite{Blasi:2009bd,Mertsch:2014poa,2012.12853,Kohri:2015mga}. 

In this work we focus on annihilation processes of the type DM DM $\rightarrow f +\bar{f}$, where $f$ indicates a specific standard model particle final state, for example the quark-antiquark mode $b \bar{b}$. 
The source term for primary CR antiprotons from DM annihilation is factorized into a spatial term and a term dependent on the antiproton kinetic energy $E_\mathrm{kin}$: 
\begin{eqnarray}
  \label{eq:pbar_DM_source_term}
  q_{\bar{p}}^{(\mathrm{DM})}(\bm{x}, E_\mathrm{kin}) =
      \frac{1}{2} \left( \frac{\rho(\bm{x})}{m_\mathrm{DM}}\right)^2  
      \sum_f \left\langle \sigma v \right\rangle_f \frac{\diff N^f_{\bar{p}}}{\diff E_\mathrm{kin}} \; ,
\end{eqnarray}
where the factor $1/2$ is for Majorana fermion DM, and $m_\mathrm{DM}$ is the DM particle mass. 
The $\rho(\bm{x})$ term in eq.~\eqref{eq:pbar_DM_source_term} indicates the DM spatial density profile. 
As done in previous works, we assume that the DM halo of our Galaxy is described by a simple NFW radial profile \cite{Navarro:1995iw} with scale radius $r_h=20$~kpc. We assign the characteristic halo density $\rho_h$ in order to reproduce a local DM density at solar position $r_\odot$ of $0.43$~GeV/cm$^3$ \cite{deSalas:2020hbh}. 
We note that the DM density is consistently rescaled among all the considered observables when computing the constraints on DM models in section~\ref{sec:ssdm}.
The sum over individual final states $f$ in eq.~\eqref{eq:pbar_DM_source_term} includes the thermally averaged annihilation cross section $ \left\langle \sigma v \right\rangle_f$  for each individual final state $f$,  and ${\diff N^f_{\bar{p}}}/{\diff E_\mathrm{kin}}$ which is the antiproton energy spectrum for a single DM annihilation.
Here we follow \cite{2107.12395} and fix the annihilation cross section independently of $f$, and assign branching fractions into different final states. We consider both a single annihilation final state ($b \bar{b}$) as well as the branching fraction structure of the scalar singlet DM model (see section~\ref{sec:ssdm} and \cite{2107.12395}). 
For both cases, we take the antiproton energy spectrum for each final state $f$ from the widely used tabulated results of \cite{Cirelli:2010xx}. These results include electroweak corrections \cite{Ciafaloni:2010ti}. For the annihilation channel into a pair of W and Z bosons, we have extended these tables to include the contribution from the off-shell production of one W or Z boson following Ref.~\cite{Cuoco:2017rxb}.

\subsubsection{Secondaries}\label{subsec:sec}

The propagation of primary CRs in the interstellar medium (ISM)  leads to the production of secondary CRs by fragmentation reactions. We recall that the ISM of our Galaxy is mainly composed of gas, with a small fraction (0.5-1\%) of dust. The interstellar gas is dominantly hydrogen (about $90$\%), and a small fraction of helium (about $10$\%). When primary CRs such as $p$, He interact with the hydrogen and helium in the ISM, secondary CR particles are produced. 
Since some CR species, such as Boron are thought to be exclusively produced through these processes, secondary-over-primary ratios of CR fluxes are powerful probes of the transport properties of our Galaxy \cite{1904.08210}. 
In the standard picture for CRs, also antiprotons are exclusively produced as secondaries \cite{Boudaud:2019efq}. 

In order to compute the source of secondary CRs we  need to convolute the flux of primary CRs ($\phi_i  ( E_{\mathrm{kin},i})$) to the ISM density ($n_{\mathrm{ISM}}$) with the energy-differential production cross section ($\frac{\diff\sigma_{ij\rightarrow \mathrm{sec}}}{\diff  E_{\mathrm{kin},\mathrm{sec}} }$): 
\begin{eqnarray}
	\label{eq:sec_source_term}
	q_{\mathrm{sec}}({\bm x},E_{\mathrm{kin},\mathrm{sec}}) &=&  
	                                \!\!\!\!\sum\limits_{j \in \lbrace \mathrm{H}, \mathrm{He} \rbrace} \!\!\!\! 4\pi \,n_{\mathrm{ISM},j}({\bm x}) 
	                                \sum\limits_{i} 
	                                \int 
	                                \diff E_{\mathrm{kin},i} \,
                                    \phi_i  ( E_{\mathrm{kin},i}) \, 
                                    \frac{\diff\sigma_{ij\rightarrow \mathrm{sec}}}{\diff  E_{\mathrm{kin},\mathrm{sec}} }(E_{\mathrm{kin},i} ,  E_{\mathrm{kin},\mathrm{sec}} )\,.
\end{eqnarray}
The source term in eq.~\ref{eq:sec_source_term} is evaluated for each CR species $q_{\mathrm{sec}}$ by considering the relevant CR primaries and the corresponding cross sections. 
The ISM density as a function of the position  in the Galaxy  follows the default \textsf{Galprop} model~\cite{Strong:1998fr}. As for the production cross section, we make different choices depending on the CR species. The secondary $p$ and He cross sections are modeled following the default \textsf{Galprop} implementation. 
As for the secondary antiprotons, we instead implement in \textsf{Galprop} the updated analytic parametrisation of the Lorentz invariant cross section as obtained in Ref.~\cite{1802.03030} as detailed in Ref.~\cite{1903.01472}.
Being tuned to all available cross section data recorded by colliders at low energies, this parametrisation implements a more reliable treatment of the production cross section for antiproton energies below about $10$~GeV. We note that for the fragmentation of primary CRs we always assume $\diff\sigma_{ij}/\diff E_{sec} = \sigma_{ij}\,\delta( E_{kin, i} - E_{kin, sec} ) $. 

Secondary CRs such as antiprotons may further scatter 
inelastically with the ISM 
and consequently lose energy. We consider this contribution, which is suppressed with respect to the secondaries, and is usually referred to as tertiary CRs \cite{Moskalenko:2001ya}.

\subsection{Propagation}\label{sec:propagation}

As outlined above, the source term of CR nuclei significantly differs between primary and secondary CRs as well as between DM and astrophysical sources. In contrast, the mechanism and, therefore, the description of CR propagation are the same for all species. The dominant process for CR nuclei, especially at high energies, is the scattering on the turbulent magnetic fields in our Galaxy. Effectively, this leads to diffusion of CRs in a halo extending a few kpc above and below the Galactic plane. From linear perturbation theory, the diffusion coefficient is expected to be antiproportional to the spectrum of magnetic wave turbulence~\cite{1306.2018}. The spectrum of magnetic turbulence depends on the exact turbulence model; the typical benchmarks are Kolmogorov or Kraichnan, both leading to power-law dependence of the diffusion coefficient as a function of rigidity, but with different spectral indices of $\delta = 0.33$ and 0.5~\cite{1995ApJ...452..912M}, respectively. 
We use a phenomenological approach and model the diffusion coefficient as a double-broken power law in rigidity
\begin{eqnarray}
    \label{eq:diffusion_coefficient}
     D_{xx} \sim  \beta R^{\delta_l}
 	 \cdot  \left( 1 + \left(\frac{R}{R_{D,0}}\right)^{1/s_{D,0}} \right)^{s_{D,0}\,( \delta - \delta_l) }  
 	 \cdot  \left( 1 + \left(\frac{R}{R_{D,1}}\right)^{1/s_{D,1}} \right)^{s_{D,1}\,( \delta_h - \delta) }\,,
\end{eqnarray}
which is normalized to $D_{xx}(R=4\;\mathrm{GV}) = D_0$. The spectral indices below, between, and above the two breaks at $R_{D,0}$ and $R_{D,1}$ are labeled $\delta_l$, $\delta$, and $\delta_h$. At the positions of the breaks, the power law is smoothed by the parameters $s_{D,0}$ and $s_{D,1}$, respectively. 
Indeed, a smooth transition is expected in the rigidity dependence of the diffusion coefficient, when  the turbulence changes between two regimes \cite{1806.04153, 1207.3706}. 
The second break at about 300~GV is directly observed in the AMS-02 data \cite{Aguilar:2021tos}. The fact that the break is more pronounced in secondaries than in primaries~\cite{Aguilar:2018njt} clearly points to a change in the diffusion coefficient rather than in the injection spectrum \cite{1706.09812}. A natural explanation for this scenario is for example provided by self-generated turbulence, as explored in Refs.~\cite{1207.3706,1806.04153}. 
In contrast, a break at lower rigidities ($\sim 10$~GV) is more speculative because also other processes like convection and reacceleration influence the CR spectra \cite{1904.08210,2103.09824}. Therefore, we will explore two different models (as in Ref.~\cite{2112.08381}), one with the low-energy break and one without. The models will be detailed further below. 

Convective winds with the velocity $\boldsymbol V_{\rm c}$ can transport the CRs away from the Galactic plane and induce adiabatic energy losses. We employ a constant convection velocity that is perpendicular to the Galactic plane, $\boldsymbol V_{\rm c} = v_{0, \rm c} \, {\rm sign}(z) \, \boldsymbol e_z$. Increasing the convection velocity decreases the CR flux below $\sim 10$~GV. In contrast, reacceleration is due to scattering of CRs on Alfv\'en magnetic waves and has the opposite effect on the CR spectra. In a head-on scattering with the Alfv\'en waves a CR gains energy while it loses energy in back-on collisions. However, head-on collisions are more likely such that statistically the CRs gain energy. In eq.~\eqref{eq:propagation_equation}, reacceleration is modeled as diffusion in momentum space with the coefficient $D_{pp}$. Larger magnetic turbulence makes reacceleration more efficient, such that $D_{pp} \sim 1/D_{xx}$~\cite{1994ApJ...431..705S}. The amount of reacceleration further depends on the Alfv\'en velocity, $D_{pp}\sim v_{\rm A}$.

The term $\partial/\partial p (\diff p / \diff t \psi_i)$ represents continuous energy losses from ionization and Coulomb collisions~\cite{astro-ph/9807150}, while the last two terms of eq.~\eqref{eq:propagation_equation} describe catastrophic losses by fragmentation and decay with the characteristic time scales $\tau_{f,i}$ and $\tau_{r,i}$, respectively.

Finally, when the CRs enter the heliosphere, they are deflected and decelerated by the solar winds. The effect is known as solar modulation and varies in a 22-year cycle. The effect is most prominent at low energies. In principle, solar modulation can be described by a diffusion equation, similar to equation \eqref{eq:propagation_equation}, but with the geometry, magnetic field and turbulence adjusted to the heliosphere. There are semi-analytical \cite{Kuhlen:2019hqb} or fully numerical codes \cite{1511.07875, 1704.03733} solving this equation numerically. We use a simplified approach instead, and treat solar modulation in the force-field approximation \cite{1976ApJ...206..333F}. 
This approximates well enough the effect of solar winds for nuclei observed above about few GeV.

\subsection{Choice of models}\label{sec:models}

We explore two distinct frameworks for CR propagation labelled INJ.BRK and DIFF.BRK. The assumptions and free parameters of each model are detailed below. 

The INJ.BRK (injection break) model has been explored extensively in literature \cite{astro-ph/9807150, astro-ph/0101231,0909.4548,1602.02243,1607.06093,2006.01337,2102.13238}. It also matches the model that we studied in the previous work on antiproton constraints~\cite{2107.12395}. The assumptions and free parameters are sketched in figure~\ref{fig:prop_models} (upper row). We employ a broken power law for the injection spectra of the primary (astrophysical) CRs with slopes $\gamma_1$ and $\gamma_2$ below and above the break position at $R_0$ with a smoothing $s$. We use different slopes for $p$ and He in the fit to account for the observed difference of the slopes in the CR fluxes of the two primaries. Understanding this difference is subject to current research. Possible explanations are, for example, different source populations \cite{1212.0381,1610.06187,1905.06699} or a $Z/A$-dependence of efficiency of Fermi-shock acceleration \cite{1704.08252,1803.00428}. The diffusion coefficient is modeled as a single broken power law with the break $R_{D,1}$ around 300 GV (\emph{i.e.} $\delta_l=\delta$ in eq.~\eqref{eq:diffusion_coefficient}). Furthermore, we allow for reacceleration and convection%
\footnote{In the recent update of \textsf{Galprop} \cite{2112.12745} from version 56 to 57 the implementation of convection was changed, correcting a false definition of the Crank–Nicolson coefficients near the Galactic plane. However, this has  negligible impact on our results because our CR fits prefer very small (compatible with 0) convection velocities.}
with $v_{\rm A}$ and $v_{0,c}$ as the two free parameters, respectively. Finally, solar modulation is treated in the force-field approximation. We allow for a slightly different solar modulation potential for antiprotons because solar modulation is charge-sign dependent \cite{AMS:2018avs}. 
\medskip

The DIFF.BRK (diffusion break) model employs a single power law for the injection spectrum of the primary CRs. This type of model has also been studied in literature (see e.g. \cite{astro-ph/9807150}), but only more recently it is tested against AMS-02 data \cite{1904.05899,2002.11406, 2103.09824}. In contrast to the INJ.BRK model, reacceleration is replaced by an additional break, $R_{D,0}$, in the diffusion coefficient, see figure \ref{fig:prop_models}. It is conceivable that the interaction of CRs and magnetic turbulence causes a damping of turbulence at low energies that effectively leads to an increase of the diffusion coefficient \cite{astro-ph/0510335}.
\medskip

For both setups, we fix the half-height of the diffusion halo to $\unit[4]{kpc}$, which is roughly the lower bound compatible with the beryllium data from AMS-02 \cite{1910.04113, 2004.00441, 2102.13238, 2103.09824, 2203.07265}. 
This is different with respect to the propagation setup used in our previous work \cite{2107.12395}, in which $z_h$ was varied as free parameter. 
However, a precise determination of the halo size is currently prevented by large systematic uncertainty in the secondary fragmentation cross section \cite{1803.04686,1910.04113, 2103.09824}. Thus, at the moment, larger values for the half-height are equally viable because of the well-known degeneracy of $z_h$ with the normalization of the diffusion coefficient. We chose a small value for the half-height because this corresponds to the most conservative DM limit. 

If we had performed the whole analysis with a different value of $z_h$, to a first approximation, we would have inferred a different value of $D_0$. The other propagation parameters would only change marginally, and also the astrophysical fluxes of primary and secondary CRs would not be affected by this. However, the DM flux would increase as a function of $z_h$. The difference between astrophysical CRs and CRs from DM annihilation is the spatial extent of the source term. The astrophysical CRs are produced in a thin layer of the Galactic plane, while DM annihilates and produces antiprotons in the entire diffusion halo~\cite{Donato:2003xg}. Empirically, we found the following enhancement factor for the flux of DM antiprotons, which is valid for $z_h$ between 1 and 10~kpc: 
\begin{eqnarray}
f_{\bar p, {\rm DM}}(z_h) &=& 
1.0
+3.7 \times 10^{-1} \left( \frac{z_h-4\,{\rm kpc}}{\rm kpc}\right) 
+5.0 \times 10^{-3} \left( \frac{z_h-4\,{\rm kpc}}{\rm kpc}\right)^2 \\ \nonumber 
&& 
-8.1 \times 10^{-3} \left( \frac{z_h-4\,{\rm kpc}}{\rm kpc}\right)^3 
+ 8.4 \times 10^{-4} \left( \frac{z_h-4\,{\rm kpc}}{\rm kpc}\right)^4  
\, .
\end{eqnarray}
We note that this factor is normalized to one at our benchmark of $z_h=4$~kpc.

\section{AMS-02 antiproton likelihood}\label{sec:ams_likelihood}

In this work, the antiproton analysis pipeline implemented using \textsf{pbarlike} obtains antiproton flux predictions from \textsf{DarkRayNet} to calculate the marginalised AMS-02 antiproton likelihood while considering correlated errors. 
Correlations for AMS-02 data have not been published and hence have to be modeled. The prescription used for modeling of data correlations and relevant theoretical uncertainties are discussed in section \ref{subsec:correlations}. \\
For the choice of propagation models DIFF.BRK and INJ.BRK, the CR propagation parameter space compatible with recent CR data was identified using \textsf{MultiNest} scans. The posterior samples from these scans were then used for training the neural networks in \textsf{DarkRayNet}. They are also used to perform marginalization over propagation and solar modulation parameters. Details about the \textsf{MultiNest} scans and marginalisation are discussed in section \ref{sec:Multinest} and section \ref{sec:marginalisation} respectively. \\
In section \ref{subsec: bb bounds}, we illustrate the effect of the new AMS-02 antiproton likelihood on constraints from the benchmark case of DM annihilation into bottom quarks.

\subsection{AMS-02 data and correlations}\label{subsec:correlations}

The likelihood $\mathcal{L}_{\bar{p}}$ for the AMS-02 measurements of the antiproton flux can be written as $-2 \log \mathcal{L}_{\bar{p}} = \chi^2_{\bar p}$, with
\begin{equation}
\label{eq:chi2corr}
 \chi^2_{\bar p}(\boldsymbol{x}_\text{DM},\boldsymbol{\theta}_\text{prop}) = \sum_{i,j} \left(\phi_{\bar{p},i}^{(\text{AMS})} - \phi_{\bar{p},i}(\boldsymbol{x}_\text{DM},\boldsymbol{\theta}_\text{prop})\right) V^{-1}_{ij} \left(\phi_{\bar{p},j}^{(\text{AMS})} - \phi_{\bar{p},j}(\boldsymbol{x}_\text{DM},\boldsymbol{\theta}_\text{prop})\right) \; ,
\end{equation}
where $\mathbf{\phi}_{\bar{p}}^{(\text{AMS})}$ denotes the AMS-02 measurements and $V$ is the covariance matrix. We estimate the covariance matrix following the approach of Refs.~\cite{1904.08210,2005.04237}, i.e.\ we write
\begin{equation}\label{eq:errors}
    V = \text{diag}(\sigma^2_{\text{stat},i}) + \sum_\alpha V^\alpha + V_\text{\rm xs} \;.
\end{equation}
The first two terms represent the uncertainty of the flux measurement by AMS-02,
where $\sigma_{\text{stat},i}$ denotes the (uncorrelated) statistical uncertainty in bin $i$ and $V^\alpha$ denote additional systematic uncertainties.\footnote{AMS-02 provides only the diagonal entries of the sum of first two terms in eq. \eqref{eq:errors}, i.e, it gives the uncorrelated statistical and systematic uncertainties in flux measurements.} A crucial contribution to the correlated uncertainty comes from the antiproton absorption cross section in the AMS-02 detector, as carefully modeled in Ref.~\cite{2005.04237}. All other contributions can be approximated by
\begin{equation}
 \left(V^\alpha\right)_{ij} = \Delta_i^\alpha \Delta_j^\alpha \exp\left(-\frac{(\log R_i/R_j)^2}{2 (l^\alpha)^2} \right) \;
\end{equation}
with $R_i$ denoting the rigidity of bin $i$, $\Delta_i^\alpha$ denoting the systematic error and $l^\alpha$ denoting the corresponding correlation length. Following Ref.~\cite{2005.04237} we consider errors in the effective acceptance, errors in the rigidity scale, unfolding errors, geomagnetic cut-off errors, template shape errors and selection errors.  

In contrast to the first two terms, the third term takes into account the uncertainties in modeling the production of secondary antiprotons. Here we follow the procedure form Refs.~\cite{1712.00002,1903.01472}. We translate the correlated uncertainties 
in the production cross section from Ref.~\cite{1802.03030} into the covariance matrix $V_{\rm xs}$ that can be applied directly on the antiproton flux. 
In more detail, we generate 2400 vectors of the cross section parameters using the correlated uncertainties from Ref.~\cite{1802.03030} and then calculate the antiproton flux with \textsf{Galprop} for each parameter vector. Finally, we have used the sample covariance for the antiproton flux evaluated at rigidities $R_i$ and $R_j$:
\begin{equation}
 \left(V_{\rm xs}\right)_{ij} = 
        \frac{1}{N-1} 
        \sum\limits^{N}_{k=1}
        \big( \phi_{\bar{p},i}^{k} - \bar \phi_{\bar{p},i} \big)
        \big( \phi_{\bar{p},j}^{k} - \bar \phi_{\bar{p},j} \big) \; ,
\end{equation}
where the superscript $k$ denotes the parameter vector and $N=2400$ is the sample size. For this procedure, the CR propagation parameters are fixed to the best-fit values, meaning that $V_\text{\rm xs}$ is slightly different for the two propagation setups DIFF.BRK and INJ.BRK. 

\subsection{Cosmic ray scans with MultiNest}
\label{sec:Multinest}

We want to gain an understanding of the reasonable parameter space of the parameters relevant for CR propagation. These will be our nuisance parameters in the likelihoods for the DM models we consider. 
We use \textsf{MultiNest} \cite{Feroz:2008xx} as a tool to efficiently sample the parameter space and obtain the likelihoods of each sampled point based on our model. \textsf{MultiNest} uses a nested sampling algorithm that initially samples from the prior and then iteratively samples and stores parameter points from ordered prior volumes according to their respective likelihoods in the parameter space. We thus get an ensemble of samples covering the entire parameter space, with the most dense sampling within the regions of high likelihood.

For each propagation model described in section~\ref{sec:propagation}, we perform this fit twice. Our initial scan includes all of the propagation parameters described previously with the assumption of no DM signal contributing to the measured CR spectra, i.e. our null hypothesis. This fit results in our most general understanding of the parameter space. We choose the priors as uniform distributions that are either constrained by observations of different CR species or have been indicated by previous parameter inferences. We simultaneously fit the most recent AMS-02 antiproton-over-proton ratio together with the fluxes of protons and helium and the $^3\mathrm{He}/^4\mathrm{He}$ ratio  \cite{Aguilar:2021tos,AMS:2019nij}. For proton and Helium we also use Voyager data \cite{Cummings:2016pdr}.  For the \textsf{MultiNest} scans, we do not consider correlations in the data sets, i.e.\ we assume that the given error bars can be interpreted as uncorrelated statistical uncertainties. The correlations in the anti-proton data, modeled as discussed above, will be included when deriving DM constraints, see section~\ref{sec:marginalisation}.

In the fit, we consider the propagation parameters going into eq.~\eqref{eq:propagation_equation} for each model. We consider the effect of solar modulation, which is modeled with the force-field potential $\varphi_\mathrm{AMS-02}$ for proton and helium. Antiprotons are allowed to have a different value of the force-field potential, as described earlier. In the fit this is taken into account by using the difference of the force-field potentials, $\varphi_\mathrm{AMS-02}-\varphi_{\bar p}$, as a free parameter. Furthermore, the normalization of the proton flux, $\mathrm{norm}_p$,%
\footnote{The global free normalization of all primary and secondary fluxes (except the DM flux) in \textsf{Galprop} is fixed by choosing the proton flux at 100\ GeV.}
and the $^4$He isotopic abundance relative to the proton abundance, ${\rm Abd}_{^{4}\mathrm{He}}$, are free parameters. Finally, we allow for cross section uncertainties in the $^3$He production cross section by introducing nuisance parameters for the cross section normalization and slope, $A_\mathrm{XS, ^{4}\mathrm{He} \rightarrow ^{3}\mathrm{He} } $ and $\delta_\mathrm{XS, ^{4}\mathrm{He} \rightarrow ^{3}\mathrm{He}} $, as in Ref.~\cite{Korsmeier:2021bkw}. The complete setup allows for independent constraints on the propagation and a fully Bayesian interpretation of all parameters.

\begin{figure}[t]
\includegraphics[height=.49\textwidth,trim={1.5cm 1.0cm 2.0cm 2.8cm},clip]{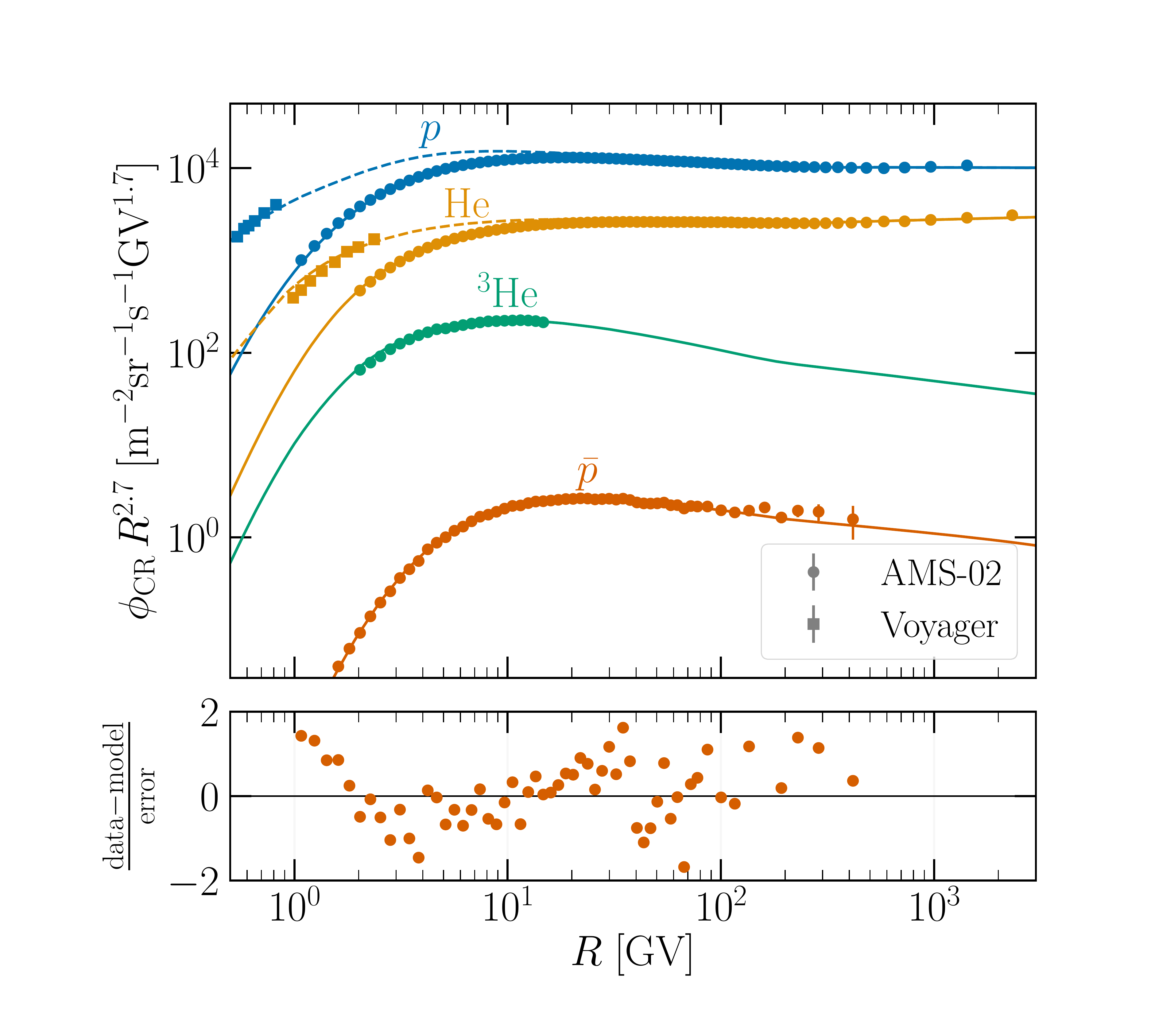}%
\includegraphics[height=.49\textwidth,trim={7.5cm 1.0cm 3.5cm 2.8cm},clip]{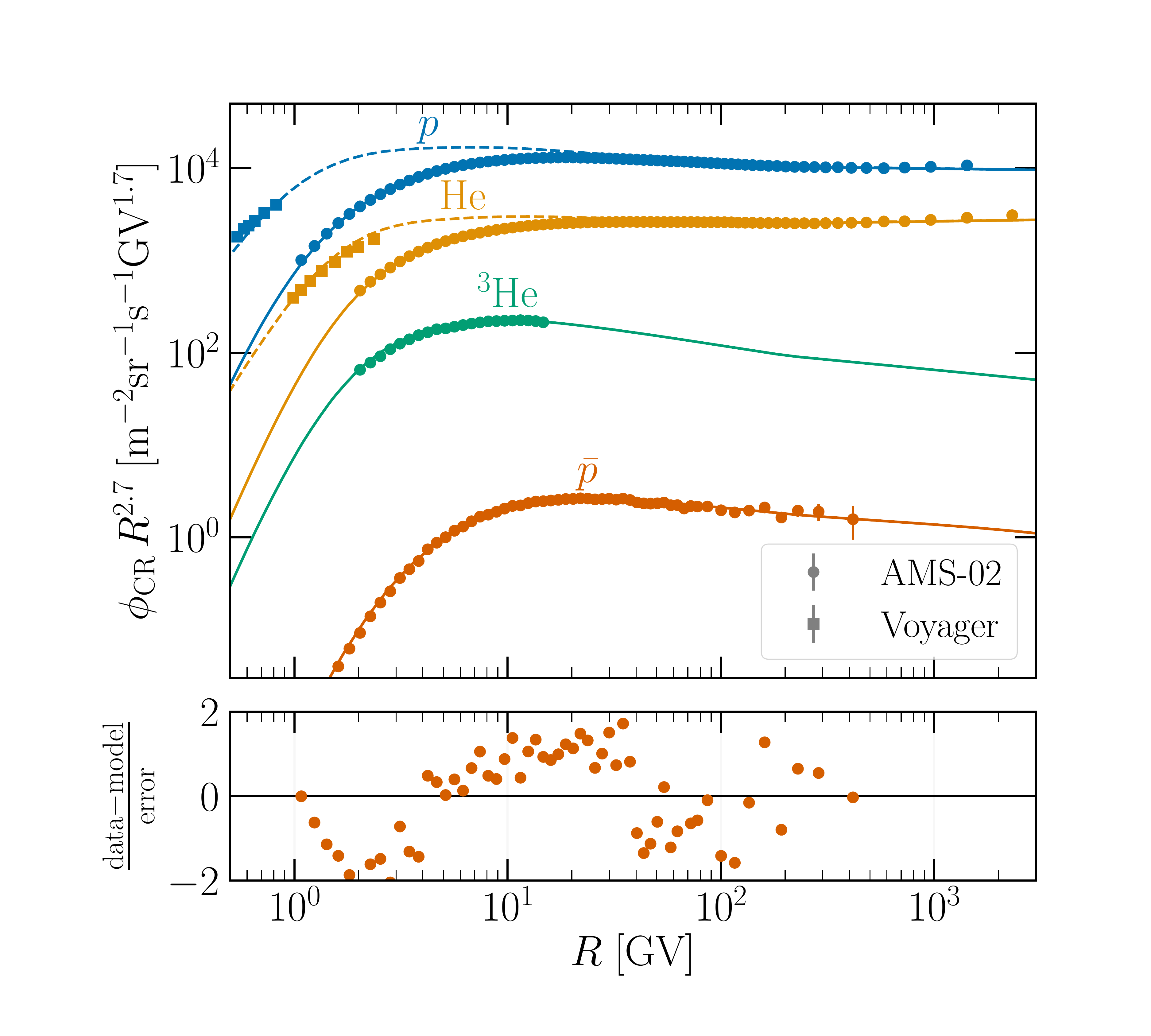}
\caption{ 
    Result of the CR fits with \textsf{MultiNest} for the two propagation setups DIFF.BRK (left) and INJ.BRK (right). We show the best fit for $p$, He, $^3$He, and $\bar{p}$. Solid lines correspond to the CR flux including solar modulation for the period of AMS-02 data-taking, while dashed lines show the local interstellar flux as measured by Voyager. In the lower panel, we show the residuals of the $\bar p$ data.
    \label{fig:Multinest_spectra}
}
\vspace{-5mm}
\end{figure}

The prior ranges and resulting best fit points of the scan (plus their 1$\sigma$ intervals) are shown in table~\ref{tab:multinest}.
We show the result of the best fit parameter point applied to the fluxes of $p, \overline{p}$, $^3$He and He for both propagation models in figure.~\ref{fig:Multinest_spectra}. While the best fit values for different physical properties vary significantly for both models, the overall fit to the data is very similar, in line with the findings of Ref.~\cite{Korsmeier:2021bkw}.  The combined $\chi^2$ here is 159.1 (163.7) for the DIFF.BRK (INJ.BRK) model. The total number of likelihood evaluations for the scan is $1.4 \, (3.7) \cdot 10^6$.

\begin{table}[t]
\caption{ 
   Priors and posterior for the CR parameters obtained from \textsf{MultiNest} fit. Results are provided for both CR propagation setups, INJ.BRK and DIFF.BRK. 
   \label{tab:multinest}
}
\centering
\renewcommand{\arraystretch}{1.3}
\begin{tabular}{lcrllcrlcrl}
\hline
\textbf{Parameters} &&  \multicolumn{3}{c}{\textbf{Priors}} && \multicolumn{2}{c}{\textbf{DIFF.BRK}} & & \multicolumn{2}{c}{\textbf{INJ.BRK}} \\
\hline
$\gamma_{1,p}$                                                       & $\,$     &                 1.2 &\!\!--&\hspace{-0.31cm} 2.1           & $\qquad$ & $               {2.338}$    &  $\!\!\!\!      ^{+0.008}_{-0.009}$   & $\,$     & $                  {1.72}$    &    $\!\!\!\! ^{+0.03}_{-0.02}$      \\
$\gamma_{1}$                                                         &          &                 1.2 &\!\!--&\hspace{-0.31cm} 2.1           &          & $               {2.280}$    &  $\!\!\!\!      ^{+0.007}_{-0.008}$   &          & $                  {1.73}$    &    $\!\!\!\! ^{+0.03}_{-0.03}$      \\
$\gamma_{2,p}$                                                       &          &                 2.1 &\!\!--&\hspace{-0.31cm} 2.6           &          & $               {2.338}$    &  $\!\!\!\!      ^{+0.008}_{-0.009}$   &          & $                 {2.448}$    &    $\!\!\!\! ^{+0.008}_{-0.007}$    \\
$\gamma_{2}$                                                         &          &                 2.1 &\!\!--&\hspace{-0.31cm} 2.6           &          & $               {2.280}$    &  $\!\!\!\!      ^{+0.007}_{-0.008}$   &          & $                 {2.393}$    &    $\!\!\!\! ^{+0.007}_{-0.007}$    \\
$R_{0}\,\mathrm{[ GV]}$                                              &          &                 1.0 &\!\!--&\hspace{-0.31cm} 20            &          & -                           &                                       &          & $                  {6.43}$    &    $\!\!\!\! ^{+0.38}_{-0.48}$      \\
$s_{}$                                                               &          &                 0.1 &\!\!--&\hspace{-0.31cm} 0.7           &          & -                           &                                       &          & $                  {0.33}$    &    $\!\!\!\! ^{+0.02}_{-0.03}$      \\
$D_{0}\,\mathrm{[ 10^{28}\;cm^2/s]}$                                 &          &                 0.5 &\!\!--&\hspace{-0.31cm} 10.0          &          & $                {3.78}$    &  $\!\!\!\!        ^{+0.10}_{-0.13}$   &          & $                  {4.10}$    &    $\!\!\!\! ^{+0.11}_{-0.08}$      \\
$\delta_{l}$                                                         &          &              $-1.0$ &\!\!--&\hspace{-0.31cm} 0.5           &          & $               {-0.66}$    &  $\!\!\!\!        ^{+0.07}_{-0.06}$   &          & $                 {0.372}$    &    $\!\!\!\! ^{+0.007}_{-0.008}$    \\
$\delta_{}$                                                          &          &                 0.3 &\!\!--&\hspace{-0.31cm} 0.7           &          & $               {0.516}$    &  $\!\!\!\!      ^{+0.010}_{-0.008}$   &          & $                 {0.372}$    &    $\!\!\!\! ^{+0.007}_{-0.008}$    \\
$\delta_h-\delta$                                                    &          &              $-0.2$ &\!\!--&\hspace{-0.31cm} 0.0           &          & $               {-0.16}$    &  $\!\!\!\!        ^{+0.02}_{-0.01}$   &          & $                 {-0.09}$    &    $\!\!\!\! ^{+0.02}_{-0.01}$      \\
$R_{D, 0}\,\mathrm{[ GV]}$                                           &          &                 1.0 &\!\!--&\hspace{-0.31cm} 20.0          &          & $                {3.91}$    &  $\!\!\!\!        ^{+0.24}_{-0.25}$   &          & -                             &                                     \\
$s_{D}$                                                              &          &                 0.1 &\!\!--&\hspace{-0.31cm} 0.9           &          & $                {0.41}$    &  $\!\!\!\!        ^{+0.02}_{-0.02}$   &          & -                             &                                     \\
$R_{D, 1}\,[ 10^{3} ] $                                              &          &                 100 &\!\!--&\hspace{-0.31cm} 500           &          & $                 {222}$    &  $\!\!\!\!        ^{+21  }_{-21  }$   &          & $                 {234  }$    &    $\!\!\!\! ^{+24  }_{-35  }$      \\
$v_{\mathrm{A}}\,\mathrm{[km/s]}$                                    &          &                   0 &\!\!--&\hspace{-0.31cm} 30            &          & -                           &                                       &          & $                 {20.40}$    &    $\!\!\!\! ^{+1.10}_{-0.64}$      \\
$v_{0,\mathrm{c}}\,\mathrm{[km/s]}$                                  &          &                   0 &\!\!--&\hspace{-0.31cm} 60            &          & $                {1.91}$    &  $\!\!\!\!        ^{+0.38}_{-1.91}$   &          & $                  {0.64}$    &    $\!\!\!\! ^{+0.10}_{-0.64}$      \\
$\varphi_{\mathrm{AMS-02}}\,\mathrm{[GV]}$                           &          &                 0.4 &\!\!--&\hspace{-0.31cm} 0.8           &          & $                {0.43}$    &  $\!\!\!\!        ^{+0.01}_{-0.01}$   &          & $                  {0.62}$    &    $\!\!\!\! ^{+0.01}_{-0.01}$      \\
$(\varphi_{\bar p} - \varphi_\mathrm{AMS-02}) \,\mathrm{[GV]}$       &          &              $-0.2$ &\!\!--&\hspace{-0.31cm} 0.2           &          & $               {0.181}$    &  $\!\!\!\!      ^{+0.019}_{-0.004}$   &          & $                {-0.165}$    &    $\!\!\!\! ^{+0.006}_{-0.035}$    \\
$\mathrm{norm}_p\,[ 10^{-8}\,\mathrm{MeV^{-1}cm^{-2}s^{-1}sr^{-1}]}$ &          &                 0.3 &\!\!--&\hspace{-0.31cm} 0.5           &          & $               {0.430}$    &  $\!\!\!\!      ^{+0.001}_{-0.001}$   &          & $                 {0.434}$    &    $\!\!\!\! ^{+0.001}_{-0.001}$    \\
${\rm Abd}_{^{4}\mathrm{He}}\,[ 10^{5} ]         $                   &          &                 0.7 &\!\!--&\hspace{-0.31cm} 1.3           &          & $               {1.063}$    &  $\!\!\!\!      ^{+0.007}_{-0.007}$   &          & $                 {0.984}$    &    $\!\!\!\! ^{+0.006}_{-0.006}$    \\
$A_\mathrm{XS, ^{4}\mathrm{He} \rightarrow ^{3}\mathrm{He} }   $     &          &                 0.8 &\!\!--&\hspace{-0.31cm} 1.2           &          & $                {1.17}$    &  $\!\!\!\!       ^{+0.01}_{-0.01}$    &          & $                 {1.177}$    &    $\!\!\!\! ^{+0.023}_{-0.006}$    \\
$\delta_\mathrm{XS, ^{4}\mathrm{He} \rightarrow ^{3}\mathrm{He}} $   &          &                -0.2 &\!\!--&\hspace{-0.31cm} 0.2           &           & $               {0.01}$    &  $\!\!\!\!       ^{+0.02}_{-0.02}$    &          & $                 {0.190}$    &    $\!\!\!\! ^{+0.010}_{-0.002}$    \\
\hline
\end{tabular} \vspace{0.4cm}\\
\renewcommand{\arraystretch}{1.2}
\end{table}

In a second step, we also include a tentative DM signal and add the DM mass and annihilation cross-section as free parameters to the \textsf{MultiNest} scan. For simplicity, we fix the annihilation channel to $b \overline{b}$, which is a standard benchmark choice. Here, we are not primarily interested in the likelihood of the final result of the scan, but rather in how the posteriors of the propagation parameters shift in order to accommodate the additional DM signal. This is important for one of the two use-cases of the \textsf{MultiNest} samples for our results. In this work we consider more general DM models than the one considered in this scan, assuming annihilation only into $b \overline{b}$, but the overall shift in the posterior for the individual propagation parameters can well be approximated with this assumption.
We have found in particular a shift for $\gamma_{1,p}$ and $\gamma_1$ at the level of about $1.7 \sigma$ towards lower values, and for $\delta$ towards a higher value at about $1.5 \sigma$ in the DIFF.BRK model. The remaining fit parameters, as well as the INJ.BRK model were affected less by the additional signal, i.e.\ the shift was at most $1 \sigma$. 

We use the combination of both scans to set up an extensive training set for the \textsf{DarkRayNet} simulation tool, a deep neural network emulator that we set up as in Ref.~\cite{2107.12395}. In the remainder of this work, this tool is used to speed up the required simulations of CR-spectra. 

For both scans in each propagation model, we store all simulated spectra with a $\Delta \chi^2 \leq$ 30 and combine them as a training set.  We include model independent DM signals by sampling four sets of DM masses and annihilation branching fractions for each set of propagation parameters. We set a flat prior on a logarithmic scale for the DM mass between 5~GeV and 5~TeV, and consider annihilation into $q \overline{q}, \ c \overline{c}, \ b \overline{b}, \ t \overline{t}, \ W^+ W^-, \ ZZ, \ gg, \ hh $, as we expect non-negligible contributions to a resulting antiproton flux from these channels, and sample random branching fractions, always summing up to one. We don't need to vary the annihilation cross section in the emulator, as it can be inserted as a normalization to the DM signal later. With the final training sample for each propagation model, the fully trained network can emulate accurate CR-spectra within an extended, relevant parameter space. With the addition of the second scan, the shift in preferred regions of the propagation parameters due to that additional signal are accounted for. 
More details on the \textsf{DarkRayNet} and the involved neural networks can be found in appendix~\ref{app:DRN}. 

The second use-case of the \textsf{MultiNest} scans is the generation of a sample of propagation parameters, which can then be used to perform the marginalization as described in more detail in the following section. For this, we extract the equally weighted posterior samples from the scan. 

\subsection{Marginalisation over propagation parameters}
\label{sec:marginalisation}

The likelihood introduced above can be used to perform a simultaneous scan over DM and propagation parameters. In the context of constraining DM models, it is however often unnecessary to infer the preferred propagation parameters. In this case, it is convenient to eliminate the dependence on the propagation parameters and obtain a likelihood function that depends exclusively on the DM parameters. Since fixing the propagation parameters to specific values may lead to overly aggressive constraints, the two possible options are either profiling or marginalisation. The former approach, i.e.\ maximising the likelihood with respect to the propagation parameters for every choice of $\boldsymbol{x}_\text{DM}$ is not only computationally expensive, but may be sensitive to finely-tuned parameter regions where small excesses in the data can be fitted. Moreover, when using \textsf{DarkRayNet} this approach may be susceptible to inaccurate network predictions for parameter regions with insufficient training (see the more detailed discussion in Ref.~\cite{2107.12395}).

The most robust approach is therefore to calculate the marginalised likelihood
\begin{equation}
\bar{\mathcal{L}}_\text{tot}(\boldsymbol{x}_\text{DM}) = \int \mathrm{d}\boldsymbol{\theta}_\text{prop} \mathcal{L}_\text{tot}(\boldsymbol{x}_\text{DM}, \boldsymbol{\theta}_\text{prop}) \pi(\boldsymbol{\theta}_\text{prop}) \; ,
\end{equation}
where $\mathcal{L}_\text{tot} = \mathcal{L}_{\bar{p}} \cdot \mathcal{L}_p \cdot \mathcal{L}_\text{He}$ and  $\pi(\boldsymbol{\theta}_\text{prop})$ denotes the assumed prior probability for the propagation parameters. In the following, we will adopt flat priors and consider the same parameter regions used for the \textsf{MultiNest} scans described above, i.e.\ we set $\pi(\boldsymbol{\theta}_\text{prop}) = \Pi_i (\theta_{i,\text{max}} - \theta_{i,\text{min}})^{-1}$. 
In principle, the marginalised likelihood can be obtained through Monte Carlo integration by drawing a random sample $\{\boldsymbol \theta_{{\rm prop}, i}\}$ of size $N$ from the prior probability distribution and writing
\begin{equation}
\bar{\mathcal{L}}_\text{tot}(\boldsymbol{x}_\text{DM}) = \frac{1}{N} \sum_{\boldsymbol \theta_{{\rm prop}, i} \sim \pi} \mathcal{L}_\text{tot}(\boldsymbol{x}_\text{DM}, \boldsymbol \theta_{{\rm prop}, i}) \; .
\label{eq:Limportance_1}
\end{equation}

In practice, a more accurate estimate\footnote{The gain in accuracy is both due to the fact that more weight is given to parameter regions where the likelihood is large, thus decreasing the uncertainty of the Monte Carlo integration, and due to the fact that \textsf{DarkRayNet} is only evaluated in parameter regions where sufficient training data is available to guarantee reliable predictions.} is obtained by drawing samples from the posterior probability for the propagation parameters in the absence of a DM signal $\mathcal{P}(\boldsymbol{\theta}_\text{prop}) \propto \mathcal{L}_\text{tot}(\boldsymbol{x}_\text{DM} = 0, \boldsymbol{\theta}_\text{prop})  \pi(\boldsymbol{\theta}_\text{prop})$. In this case,
\begin{equation}
\bar{\mathcal{L}}_\text{tot}(\boldsymbol{x}_\text{DM}) \propto \frac{1}{N} \sum_{\boldsymbol \theta_{{\rm prop}, i} \sim \mathcal{P}} \frac{\mathcal{L}_\text{tot}(\boldsymbol{x}_\text{DM}, \boldsymbol \theta_{{\rm prop}, i})}{\mathcal{L}_\text{tot}(\boldsymbol{x}_\text{DM} = 0, \boldsymbol \theta_{{\rm prop}, i})} \; .
\label{eq:Limportance_2}
\end{equation}
The constant of proportionality is independent of $\boldsymbol{x}_\text{DM}$ and therefore drops out when calculating likelihood ratios. 

Now we can make use of the fact that the contribution of DM to the local proton and Helium flux is completely negligible. Hence, the proton and Helium likelihoods $\mathcal{L}_{p,\text{He}}$ are independent of $\boldsymbol{x}_\text{DM}$ and therefore cancel in the likelihood ratio. We can therefore replace $\mathcal{L}_\text{tot} \to \mathcal{L}_{\bar{p}}$ in the equation above. The marginalised $\chi^2$ is then given by
\begin{align}
\bar{\chi}^2(\boldsymbol{x}_\text{DM}) & = -2 \log\left( \sum_{\boldsymbol \theta_{{\rm prop}, i} \sim \mathcal{P}} \frac{\mathcal{L}_{\bar{p}}(\boldsymbol{x}_\text{DM}, \boldsymbol \theta_{{\rm prop}, i})}{\mathcal{L}_{\bar{p}}(\boldsymbol{x}_\text{DM} = 0, \boldsymbol \theta_{{\rm prop}, i})} \right) \nonumber \\
& = -2 \log\left[ \sum_{\boldsymbol \theta_{{\rm prop}, i} \sim \mathcal{P}} \exp \left( -\frac{ \chi^2_{\bar p}(\boldsymbol{x}_\text{DM}, \boldsymbol \theta_{{\rm prop}, i}) - \chi^2_{\bar p}(\boldsymbol{x}_\text{DM} = 0, \boldsymbol \theta_{{\rm prop}, i})}{2} \right) \right]
\label{eq:Limportance_3}
\end{align}
with $\chi^2_{\bar p}$ defined in eq.~\eqref{eq:chi2corr}.
We note that by construction $\bar{\chi}^2(\boldsymbol{x}_\text{DM}=0) = 0 $.

In general, drawing a random sample from the posterior probability is far from easy. Fortunately, this sample only needs to be generated once and can then be used for arbitrary DM parameters. Moreover, such a sample is generated automatically as a by-product of the \textsf{MultiNest} scans that we have run to generate the training data for \textsf{DarkRayNet} and therefore requires no additional calculations. This sample of about $10^4$ sets of propagation parameters is included in \textsf{pbarlike} and used by default for marginalisation. 

To conclude this discussion, we note that it is in fact possible to use a different definition of the antiproton likelihood for the marginalised likelihood than the one that has been used to calculate the posterior probability. In this case, eq.~\eqref{eq:Limportance_3} simply becomes 
\begin{equation}
\bar{\mathcal{L}}_\text{tot}(\boldsymbol{x}_\text{DM}) \propto \frac{1}{N} \sum_{\boldsymbol \theta_{{\rm prop}, i} \sim \mathcal{P}_0} \frac{\mathcal{L}_\text{tot}(\boldsymbol{x}_\text{DM}, \boldsymbol \theta_{{\rm prop}, i})}{\mathcal{L}_{\text{tot},0}(\boldsymbol{x}_\text{DM} = 0, \boldsymbol \theta_{{\rm prop}, i})}
\end{equation}
with 
$\mathcal{P}_0(\boldsymbol{\theta}_\text{prop}) \propto \mathcal{L}_{\text{tot},0}(\boldsymbol{x}_\text{DM} = 0, \boldsymbol{\theta}_\text{prop})  \pi(\boldsymbol{\theta}_\text{prop})$. This means in particular that it is possible to modify the covariance matrix without the need to rerun the computationally expensive \textsf{MultiNest} scan.

\subsection{Constraints on DM annihilations into $b\bar{b}$} \label{subsec: bb bounds}
Assuming all of DM annihilates only into bottom quarks, we calculate the marginalised log-likelihood ratios \textsf{pbarlike}, for each point on a 2D grid of DM model parameters $\{m_{DM}, \langle \sigma v \rangle \}$. We consider the DM parameter ranges, $m_{DM} \in [10 \ \unit{GeV}, 5 \ \unit{TeV}]$ and $\langle\sigma v\rangle \in [10^{-28}, 10^{-24}] \ \unit{cm^{3}s^{-1}}$. An entire routine consisting of obtaining antiproton flux prediction from \textsf{DarkRayNet}, solar modulation and marginalised log-likelihood ratio calculation for a single point in DM model parameter space takes $\mathcal{O}(10) \ \unit{s}$. The evaluation of $\mathcal{O}(10^4)$ antiproton fluxes (for the entire sample of  $\mathcal{O}(10^4)$ propagation parameter vectors used for marginalization) using \textsf{DarkRayNet} takes only $\mathcal{O}(1) \, \unit{s}$, whereas \textsf{Galprop} would take $\mathcal{O}(10)$ \, \unit{s} for each set of propagation parameters. The most time-consuming part of the likelihood calculation turns out to be the modulation of each of $\mathcal{O}(10^4)$ antiproton fluxes with appropriate solar modulation parameters.

\begin{figure}[t]
\includegraphics[width=\textwidth]{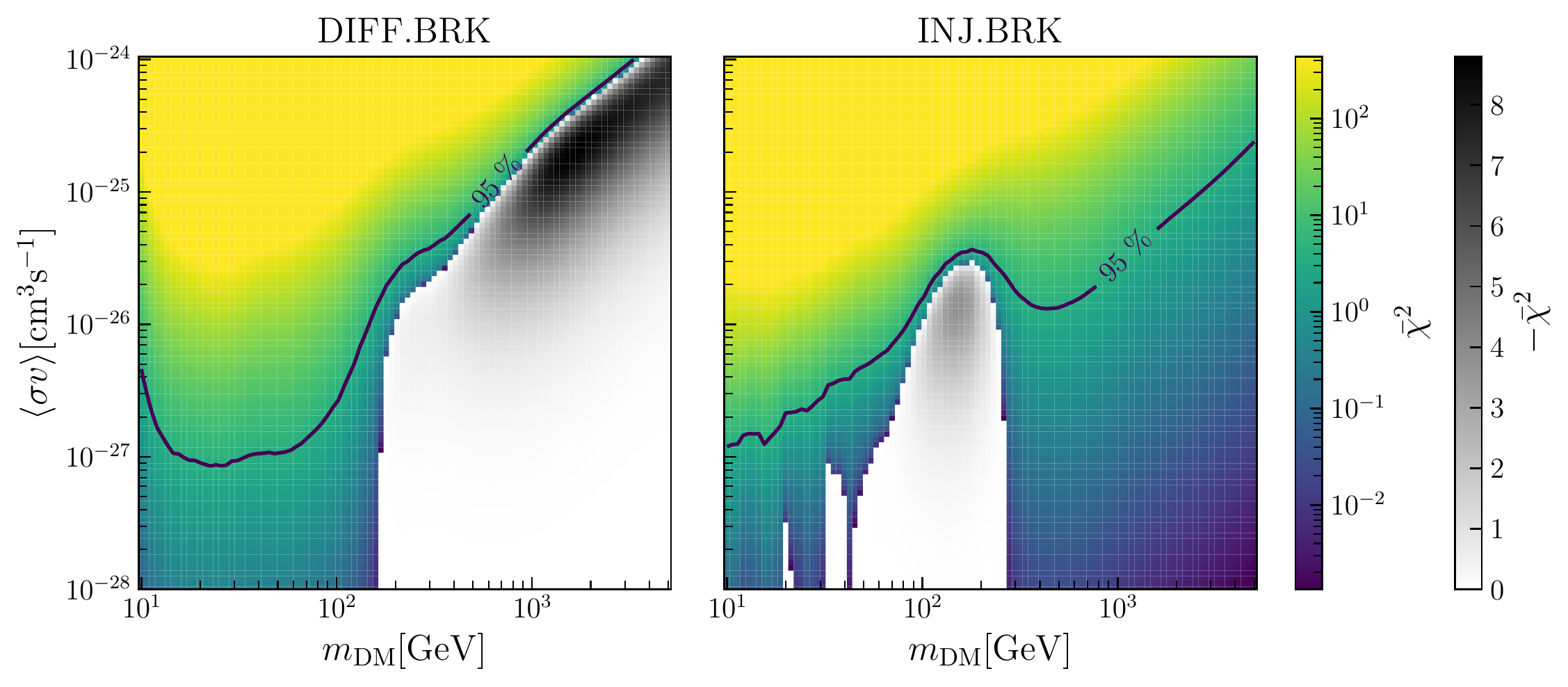}
		\caption{ 
            $\Bar{\chi}^2$ for DM annihilating into $b \Bar{b}$, obtained using the two CR propagation models-DIFF.BRK (\textit{left}) and INJ.BRK (\textit{right}). The black line represents the $95\%$ CL upper limit for $\langle{\sigma v}\rangle$ as a function of $m_{DM}$. 
        }
		\label{fig:bb bounds}
\end{figure}

The results are shown in figure \ref{fig:bb bounds}. The region in colour correspond to $\Bar{\chi}^2 > 0$ and hence prefers the case of antiprotons from a purely secondary origin. The region in gray corresponds to $\Bar{\chi}^2 < 0$ and  thus exhibits a small preference for the presence of a DM signal. The black line shows the $95\%$ CL upper bound for $\langle\sigma v\rangle$ as a function of DM mass $m_{DM}$ (corresponding to $\Bar{\chi}^2 = 3.84$). For the DIFF.BRK propagation model (left panel), preference for DM signal is seen in $\unit{TeV}$ masses and for INJ.BRK (right panel) around $\unit[100]{GeV}$. The bound obtained using INJ.BRK model is similar to ones previously seen in literature~\cite{1712.00002,1903.01472,2005.04237,Calore:2022stf}. For the DIFF.BRK model, however, the DM preference is pushed to $\unit{TeV}$ masses. This can be understood with the help of residuals in figure \ref{fig:Multinest_spectra}. The additional break at low rigidities in the diffusion coefficient in the DIFF.BRK model allows for a good fit to low rigidity data, whereas it performs a poorer fit to data at high rigidities compared to INJ.BRK model.

To conclude this section, we note that \textsf{pbarlike} also returns log-likelihood ratios calculated using uncorrelated errors given by AMS-02. For more details, we refer to appendix~\ref{app:uncorr}.

\section{Application to scalar singlet dark matter}\label{sec:ssdm}

We are now in the position to apply the AMS-02 antiproton likelihood to more realistic DM models. As an example, we consider the scalar singlet DM model, which is briefly reviewed in section~\ref{sec:details}. We discuss the additional constraints that we have implemented in section~\ref{sec:constraints} and present our results in section~\ref{sec:results}.

\subsection{Model details}
\label{sec:details}

In the scalar singlet DM model, the SM is extended by a gauge-singlet real scalar boson $s$ stabilised by a $\mathbb{Z}_2$ symmetry~\cite{Silveira:1985rk,McDonald:1993ex,Burgess:2000yq}. The resulting scalar potential can then be written as
\begin{equation}
V(s^2,H^{\dagger}H) = \lambda_h \left[\left(H^{\dagger}H\right) - \frac{v^2}{2}\right]^2 + \frac{1}{2} \, \lambda_{hs} \, s^2 \, H^{\dagger}H + \frac{1}{4} \, \lambda_s \, s^4 + \frac{1}{2} \, m_{s,0}^2 \, s^2
\end{equation}
with $v = 246 \: \text{GeV}$ denoting the vacuum expectation value of the SM Higgs field. After electroweak symmetry breaking, we can replace $H = (h + v, 0)/\sqrt{2}$ with a real scalar $h$ to obtain the potential
\begin{equation}
 V(s^2,h) = V(h) + \frac{1}{2} \, m_{s}^2 \, s^2 + \frac{1}{4} \, \lambda_s \, s^4 + \frac{1}{2} \, \lambda_{hs} \, v \, h \, s^2 + \frac{1}{4} \, \lambda_{hs} \, h^2 \, s^2 \; ,
\end{equation}
where $m_{s} = \left[m_{s,0}^2 + \lambda_{hs} \, v^2 / 2\right]^{1/2}$ denotes the physical mass of the scalar singlet.

The quartic self-coupling $\lambda$ typically has no consequences for phenomenology, and we will therefore set it to zero in the following. The model is then fully characterised by the scalar singlet mass $m_s$ and the Higgs portal coupling $\lambda_{hs}$.\footnote{In the analysis below, we also include a number of nuisance parameters, most notably the local DM density $\rho_\odot$, which affects direct and indirect detection constraints. See Ref.~\cite{1705.07931} for details.} The latter is responsible for all interactions of $s$ with SM particles, which determine the relic density of $s$ via the freeze-out mechanism and the potentially observable signatures of $s$ in satellites and laboratory experiments. 

\subsection{Implementation of new constraints}
\label{sec:constraints}

The phenomenology of the scalar singlet DM model has been discussed in great detail in Refs.~\cite{1306.4710,1705.07931,1806.11281}. Here, we repeat the analysis procedure from Ref.~\cite{1705.07931}, i.e.\ we calculate the relic density $\Omega_s$ of scalar singlets and require $f_s \equiv \Omega_s / \Omega_\text{DM} \leq 1$, where $\Omega_\text{DM} = 0.12$~\cite{Planck:2018vyg}. Furthermore, we consider constraints from direct detection experiments (rescaled by the fractional abundance $f_s$ of scalar singlets), indirect detection experiments (rescaled by a factor $f_s^2$) and the LHC as well as the perturbativity requirement $\lambda_s < \sqrt{4\pi}$. In addition to previously considered constraints, we include the antiproton likelihood from AMS-02 discussed above, as well as a number of additional new likelihoods as detailed below.

\subsubsection{Higgs invisible width} 

The partial width for the decay of a Higgs boson into a pair of scalar singlets is given by
\begin{equation}
\Gamma(h \rightarrow ss) = \frac{\lambda_{hs}^2 v^2}{32\pi \, m_h} \sqrt{1 - \frac{4 \, m_s^2}{m_h^2}} \; ,
\end{equation}
from which the invisible branching ratio can be calculated as
\begin{equation}
\text{BR}(h \rightarrow ss) = \frac{\Gamma(h \rightarrow ss)}{\Gamma(h \rightarrow ss) + \Gamma_\text{SM}} \; .
\end{equation}
Both CMS~\cite{2201.11585} and ATLAS~\cite{2202.07953} have recently published new constraints on the Higgs invisible branching ratio using the vector boson fusion channel.

\textbf{CMS.} The CMS constraint is provided in the form of a likelihood as a function of $\text{BR}(h \to ss)$, see figure 12 of Ref.~\cite{2201.11585}. This likelihood can be very well approximated by the quadratic function
\begin{equation}
-2 \log \mathcal{L} = a (\text{BR}(h \to ss) - b)^2
\label{eq:Hinvlike}
\end{equation}
with $a \approx 339$ and $b \approx 0.089$.

\textbf{ATLAS.} For ATLAS, no detailed likelihood function is available. However, assuming that the ATLAS likelihood can also be written in the form of eq.~\eqref{eq:Hinvlike}, we can infer the two parameters from the stated bounds $\text{BR}(h \to ss) < 0.127 \ (0.145)$ at 90\% (95\%) confidence level, which yields $a \approx 303$ and $b \approx 0.032$.

The combined constraint from CMS and ATLAS gives $\text{BR}(h \to ss) < 0.14$ at 95\% confidence level. The best-fit value is $\text{BR}(h \to ss) = 0.06$, which is preferred over $\text{BR} = 0$ with a significance of $1.2\sigma$.

\subsubsection{Direct detection}

The DM--nucleon scattering cross section at zero momentum transfer is given by
\begin{equation}
 \sigma_\text{SI} = \frac{\lambda_{hs}^2 f_N^2}{4\pi} \frac{\mu^2 \, m_N^2}{m_h^4 \, m_s^2} \; ,
\end{equation}
where $m_N$ is the nucleon mass, $\mu = (m_s \, m_N) / (m_s + m_N)$ is the reduced mass and $f_N$ is the effective Higgs-nucleon coupling. Recent constraints on $\sigma_\text{SI}$ as a function of $m_s$ have been published by LZ and PandaX.

\textbf{LZ.} We consider events below the 90\% quantile of the nuclear recoil band, such that the effective exposure is $2.97 \cdot 10^5 \, \mathrm{kg \, days}$. At $E_\mathrm{R} = 15 \,\mathrm{keV}$ we split the search region into two bins, assuming an energy resolution of $\sigma = 1.5\,\mathrm{keV}$. We then fix the background expectation in each bin in such a way that the expected limit is recovered when setting the observed number of events equal to the background expectation. We find that this procedure yields 1 (7) expected background event in the lower (upper) bin. The actual observation gave no events in the lower and twelve events in the upper bin. This leads to an observed exclusion limit somewhat stronger (slightly weaker) than the expected one for small (large) DM masses, in agreement with the published LZ result.

\textbf{PandaX-4T.} 
We consider events below the mean of the nuclear recoil band, leading to an effective exposure of $1.15 \cdot 10^5 \, \mathrm{kg \, days}$. In this region, the expected background was 9.8 events, compared to 6 observed events. To calculate the signal prediction, we take the efficiency from figure 2 of Ref.~\cite{2107.13438}.

\subsection{Results}
\label{sec:results}

In this section, we present updated constraints on the scalar singlet DM model. To do this, one could perform a comprehensive global scan with all the relevant available datasets and the resulting $\mathcal{O}(10)$ nuisance parameters. Instead, for a judicious use of computational resources, we take advantage of the publicly available results \cite{supp1705.07931} (hereafter referred to as GC17) from the extensive global analysis performed by the GAMBIT collaboration in Ref.~\cite{1705.07931}. This analysis combined the then available relic density measurements, direct detection limits, limits from invisible Higgs decays, and indirect detection limits from dark annihilating into neutrinos and gamma-rays. The results in GC17 contain parameter spaces allowed by the limits considered and their corresponding combined likelihoods. 

The large  sample of $\mathcal{O}(10^7)$ points in GC17 contains combined results from multiple sampling runs. These scans included the DM model parameters in the range $m_s \in [\unit[45]{GeV},\unit[10]{TeV}]$ and $\lambda_{hs} \in [10^{-4},10]$. For details on the nuisance parameters and their ranges, see table 2 of Ref.~\cite{1705.07931}. The scans identified the best-fit at $\lambda_{hs} = 6.5 \times 10^{-4}, \ m_s = \unit[62.51]{GeV}$ corresponding to $f_s^2 \langle\sigma v\rangle_0 = \unit[8.65 \times 10^{-28}]{cm^3 s^{-1}} $.

\begin{figure}[t]
    \centering
    \includegraphics[height=0.38\linewidth,clip,trim=0 45 60 0]{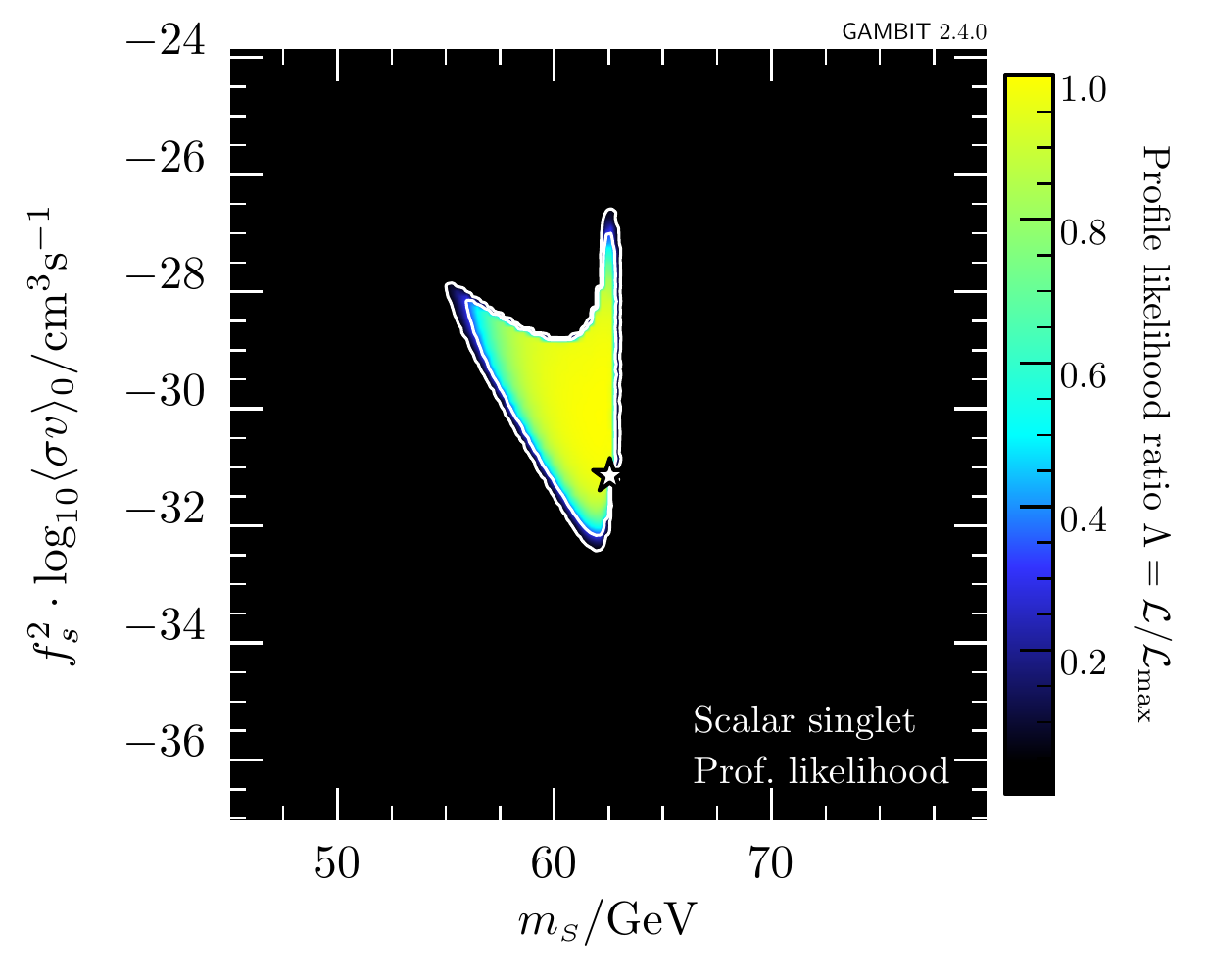} 
    \includegraphics[height=0.38\linewidth,clip,trim=60 45 0 0]{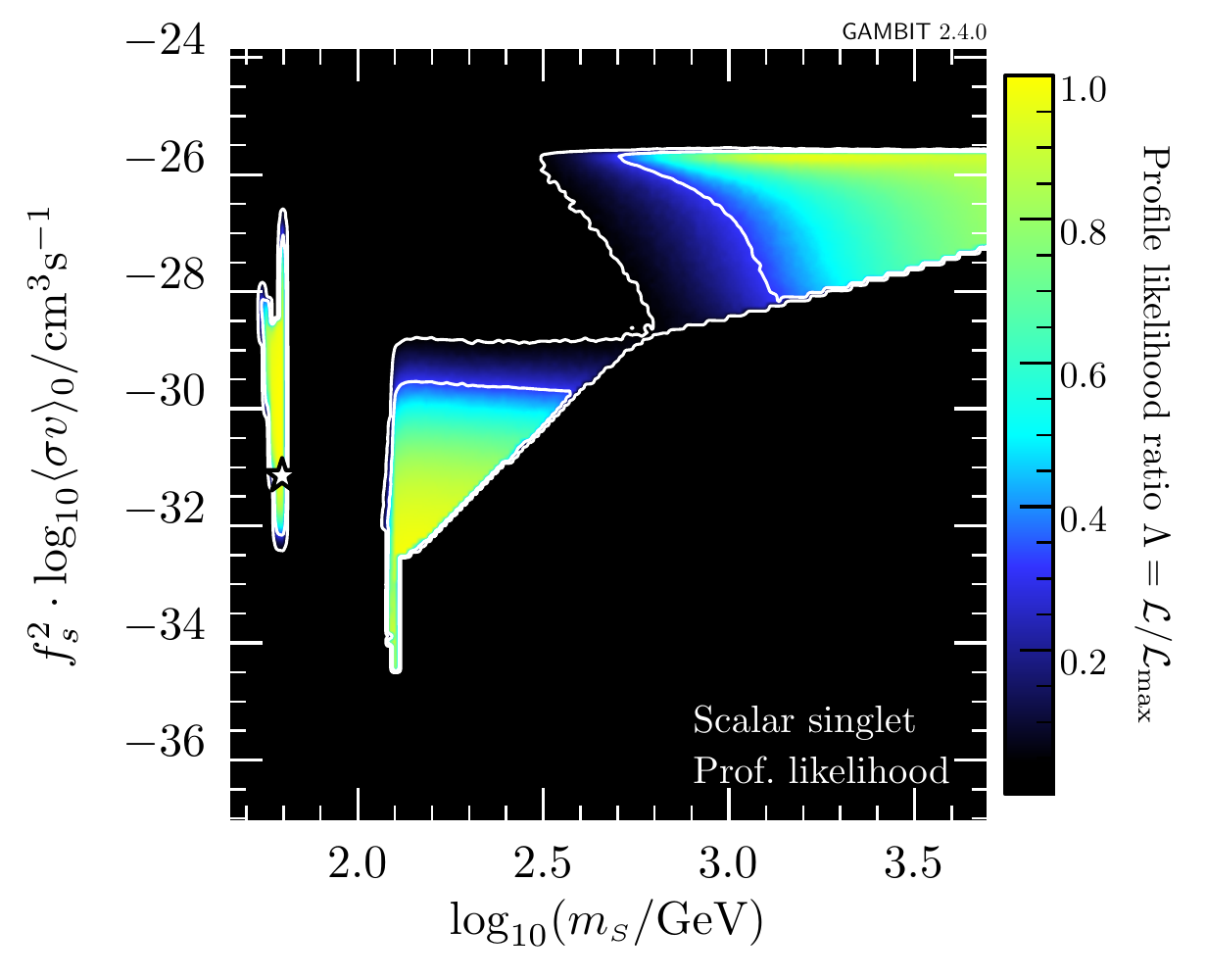} \\
    \includegraphics[height=0.45\linewidth,clip,trim=0 0 60 0]{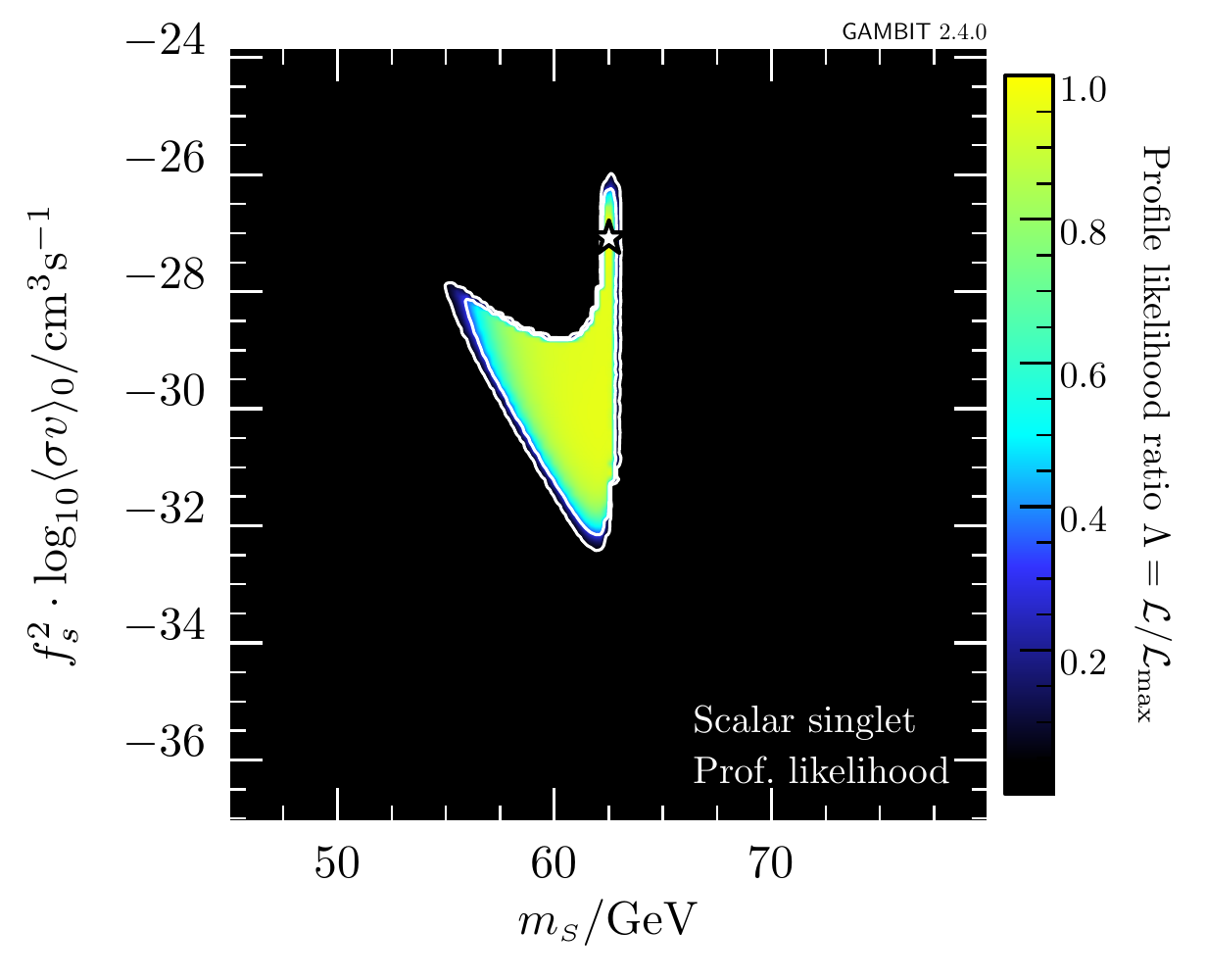}
    \includegraphics[height=0.45\linewidth,clip,trim=60 0 0 0]{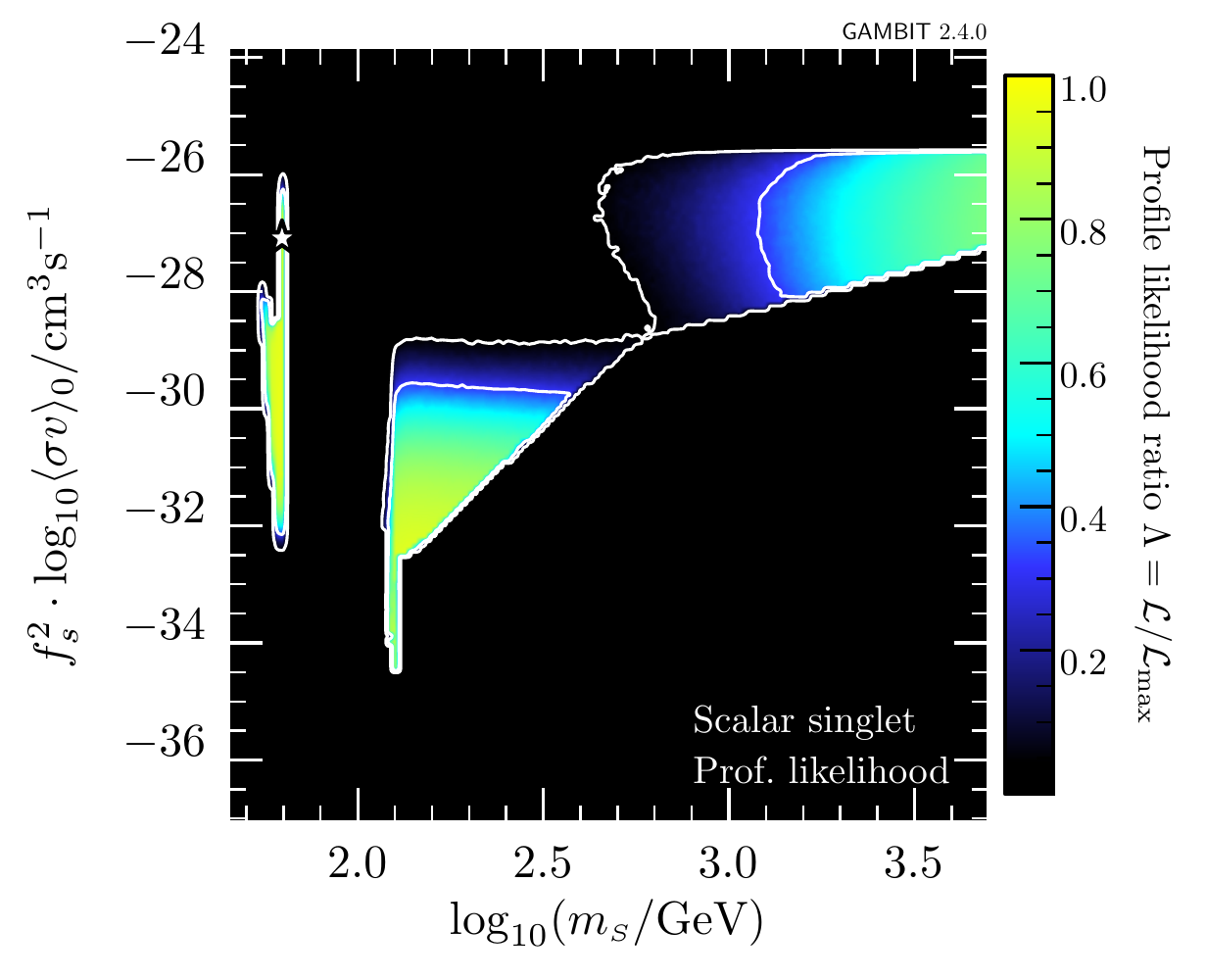}  
    \caption{Allowed parameter space of the scalar singlet model in the $f_s^2 \langle\sigma v \rangle_0$--$m_s$ plane. The plots show the modification of parameter space, on updating constraints in \cite{1705.07931} with the addition of AMS-02 antiproton likelihood. The top (bottom) row shows constraints obtained using DIFF.BRK (INJ.BRK) model.
    \label{fig:ssdm_sigmav}}
\end{figure}

For our analysis, we restrict the parameter space to $m_s \in [\unit[45]{GeV},\unit[5]{TeV}]$ to remain within the parameter region where \textsf{DarkRayNet} has been trained. We further reduce the GC17 sample to $\mathcal{O}(10^5)$ points. We then update the old combined likelihoods in GC17 by adding the new likelihoods discussed above using the \texttt{postprocessor} scanner available within the \textsf{GAMBIT} sampling module \texttt{ScannerBit} \cite{scannerbit_2017}.  We emphasize that this procedure is possible because there is no strong DM signal in the AMS-02 data, and hence we do not expect the new likelihoods to open up previously disfavoured parameter regions.

In a first step, we only add the new AMS-02 antiproton likelihood, to explore its impact on the allowed parameter space. The resulting updated constraints on the $f_s^2 \langle \sigma v \rangle_0$--$m_s$ plane are shown in figure \ref{fig:ssdm_sigmav}.\footnote{Plots were made using \textsf{pippi}~\cite{scott_pippi_2012}.} Each panel shows the profiled likelihood in the DM model parameter space, normalized by the best-fit likelihood.\footnote{Note that the total likelihood here is profiled over the nuisance parameters in GC17 only; \textsf{pbarlike} already returns the marginalized likelihood and hence, the CR propagation and solar modulation parameters are not included in the \textsf{GAMBIT} postprocessing run as nuisance parameters.}  The contours show the $1\sigma$ and $2\sigma$ confidence regions, and the star indicates the best-fit point. The viable regions identified in previous studies, i.e.\ the resonance region with $m_s \approx m_h/2$ and small cross sections (left column) and the high-mass regions with $m_s \gg m_h$ and large cross sections (right column), are still retained. For comparison, see the right panel of figure 3 in Ref.~\cite{1705.07931}.

The two rows of figure~\ref{fig:ssdm_sigmav} correspond to the DIFF.BRK (top) and INJ.BRK (bottom) models.  For both propagation models, the likelihood of the best-fit point is changed by $-2 \Delta \log \mathcal{L} < 0.1$. Neither propagation models shift the best-fit point to different masses. For DIFF.BRK, the best-fit point is however shifted to smaller couplings and a smaller log-likelihood, as this model leads to strong bounds on the rescaled annihilation cross section for DM masses in the 10--100$\,\mathrm{GeV}$ range (see also figure~\ref{fig:bb bounds}). For the INJ.BRK model, the small excess seen in figure~\ref{fig:bb bounds} leads to a weaker bound on the rescaled cross section and a correspondingly larger coupling at the best-fit point with a larger log-likelihood. In the TeV range, on the other hand, the INJ.BRK model gives the stronger constraint than the DIFF.BRK model, leading to a smaller allowed parameter region. This observation highlights the relevant impact of the AMS-02 antiproton constraints on the scalar singlet DM model.

\begin{figure}[t]
    \centering
    \includegraphics[height=0.38\linewidth,clip,trim=0 45 60 0]{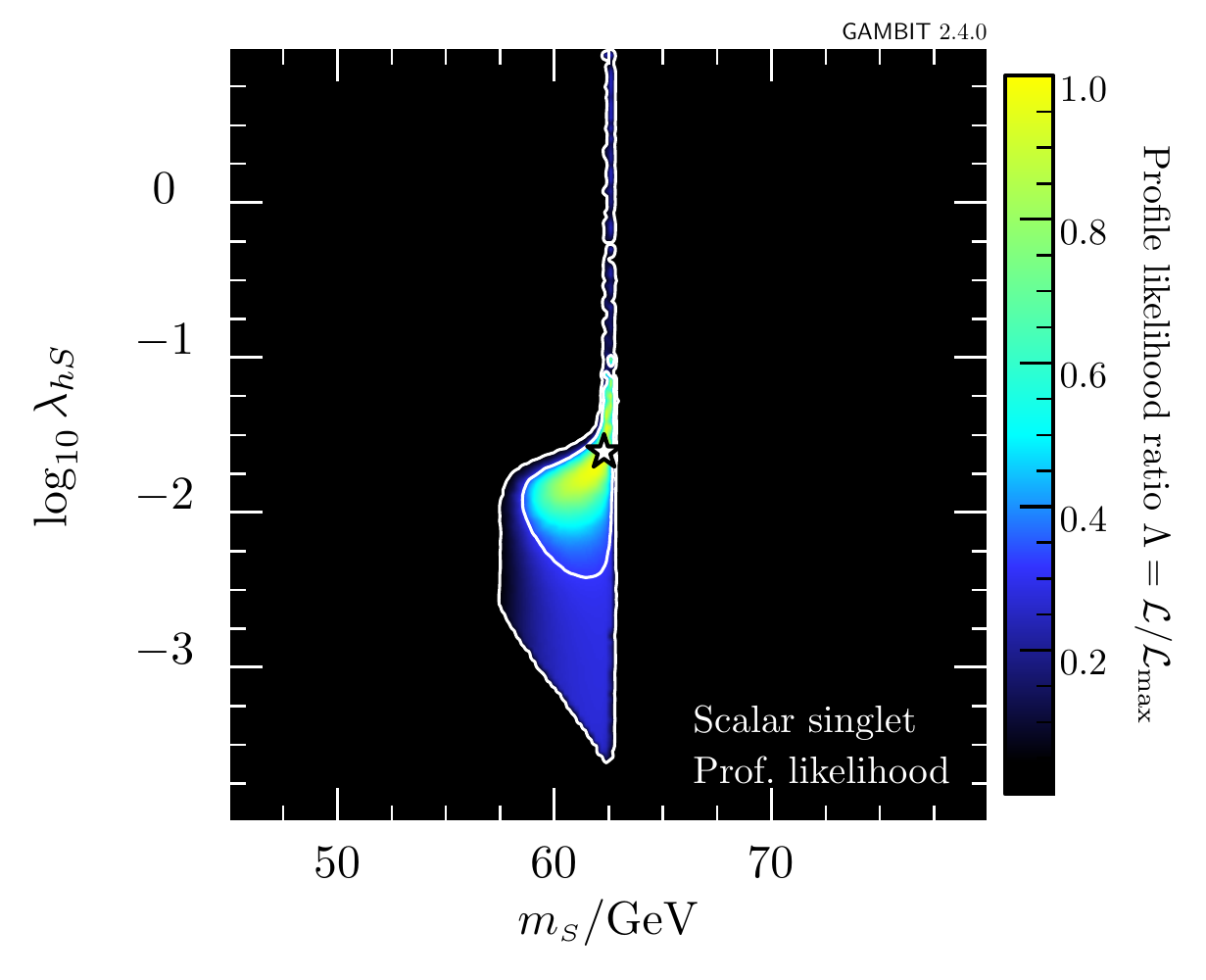} 
    \includegraphics[height=0.38\linewidth,clip,trim=60 45 0 0]{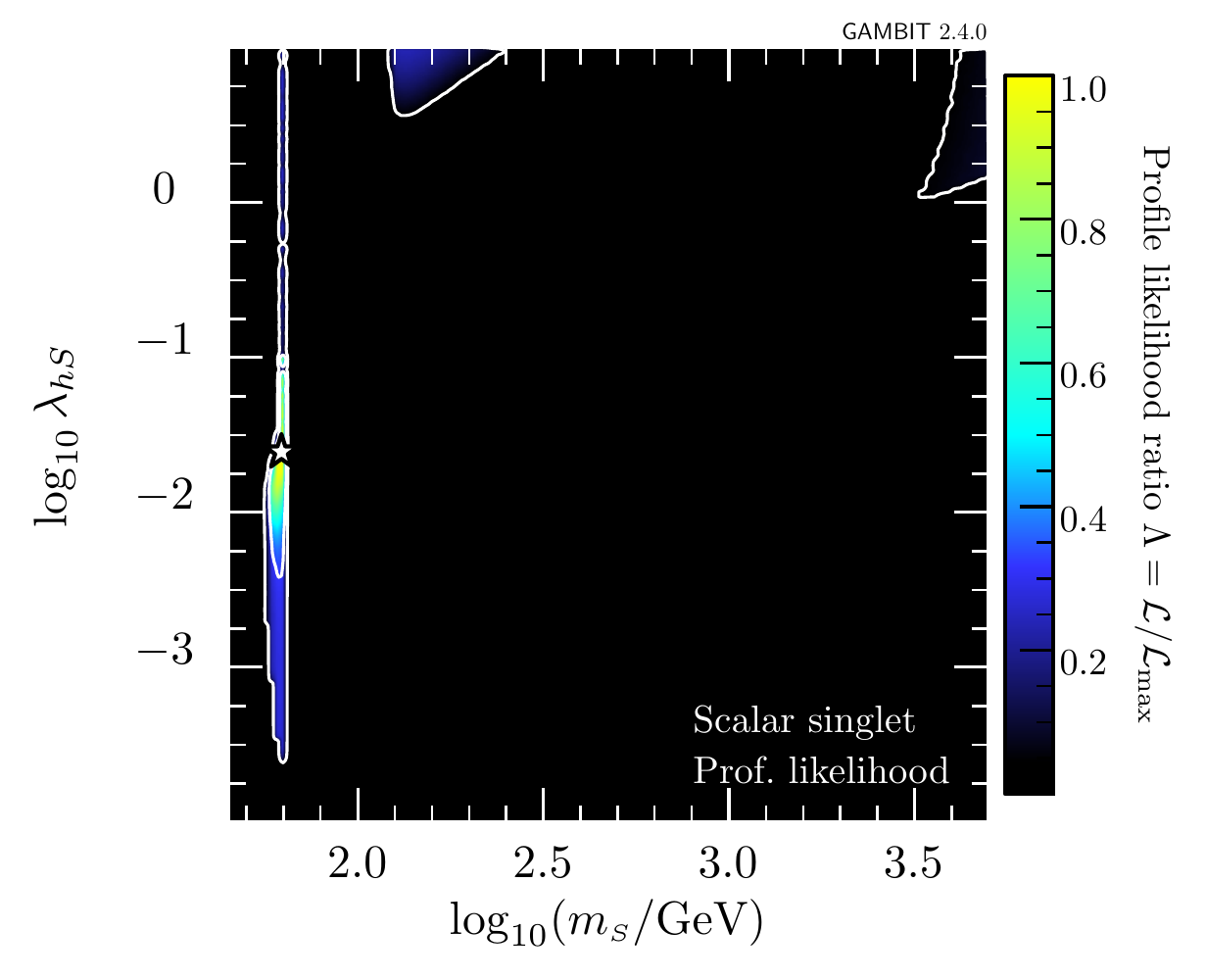} \\
    \includegraphics[height=0.45\linewidth,clip,trim=0 0 60 0]{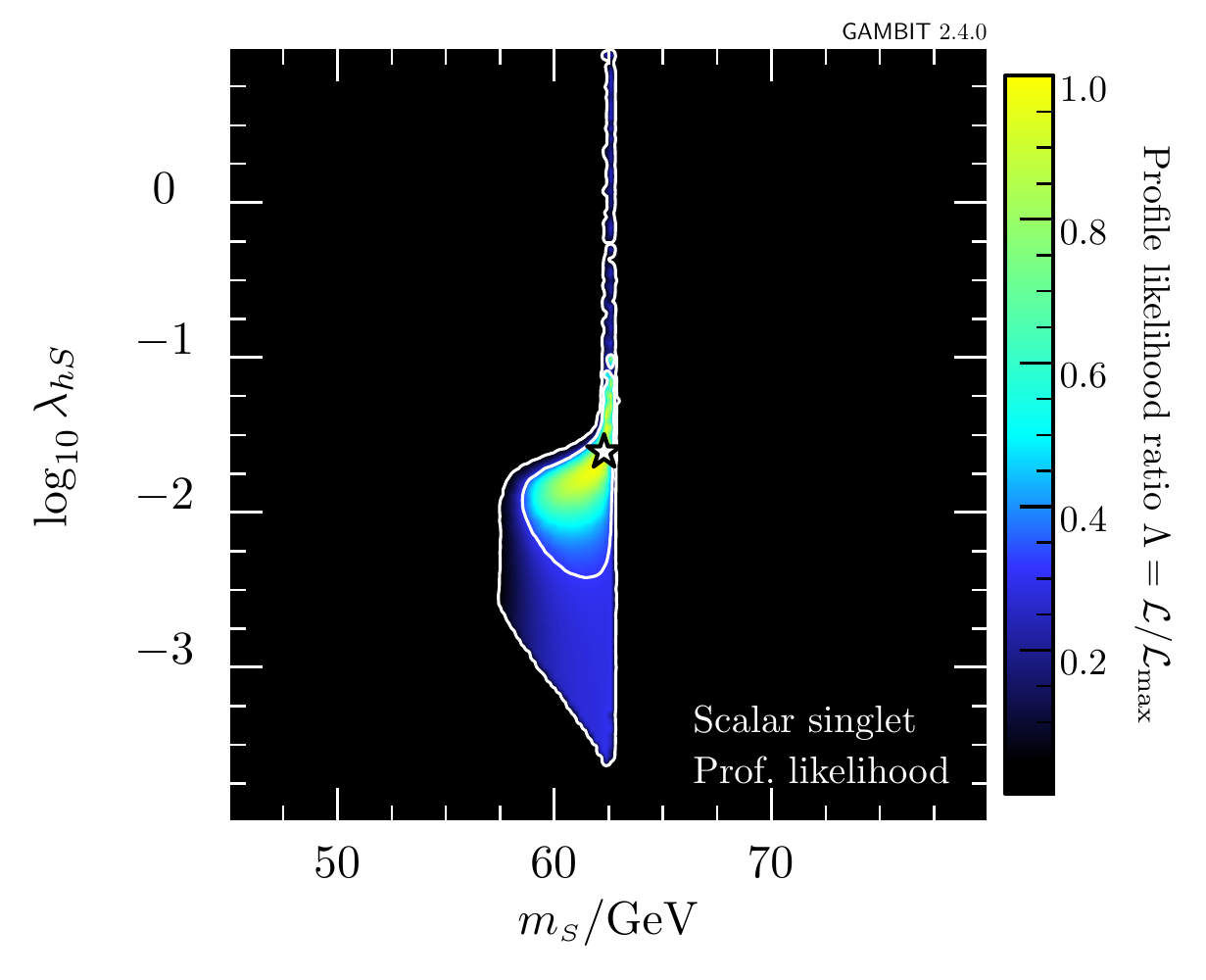}
    \includegraphics[height=0.45\linewidth,clip,trim=60 0 0 0]{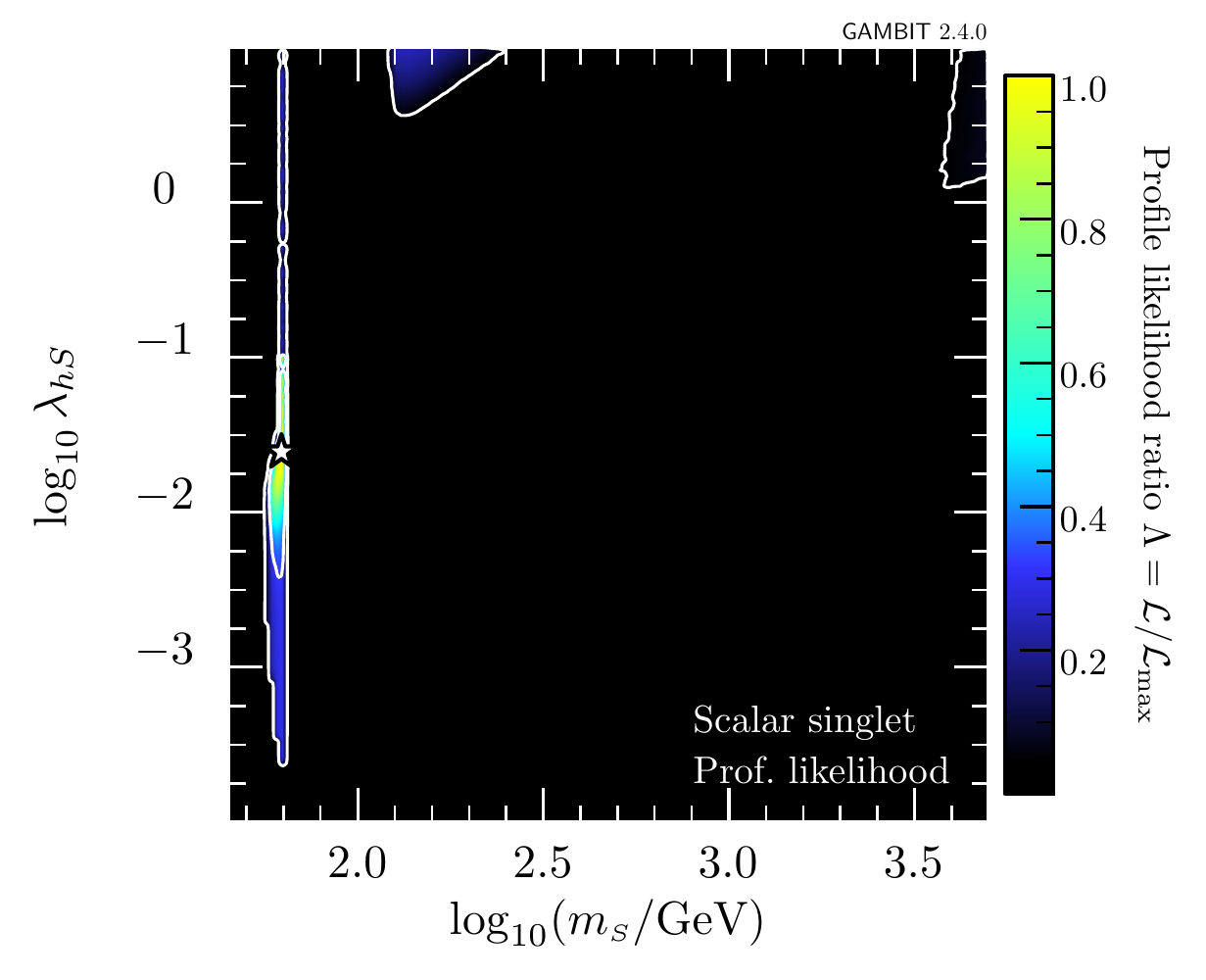}  
    \caption{Allowed parameter space of the scalar singlet model in the $\lambda_s$--$m_s$ plane. In addition to the constraints considered in figure~\ref{fig:ssdm_sigmav}, we also include the most recent constraints from direct detection experiments (PandaX-4T and LZ) and the most up-to-date measurements of the Higgs invisible branching ratio from ATLAS and CMS. In the top (bottom) row, we consider the AMS-02 antiproton likelihood using the DIFF.BRK (INJ.BRK+vA) propagation model. 
    \label{fig:ssdm_lambdas}}
\end{figure}

In a second step, we now also include the new direct detection constraints using \textsf{DDCalc}~\cite{GAMBITDarkMatterWorkgroup:2017fax,GAMBIT:2018eea} and the Higgs invisible branching ratio likelihood, which has been directly implemented within \textsf{GAMBIT}. The resulting updated constraints on the $\lambda_{hs}$--$m_s$ plane are shown in figure \ref{fig:ssdm_lambdas} for the DIFF.BRK (top) and INJ.BRK (bottom) models. As before, the figure also shows a zoom-in of the resonance region with mass in linear scale (left). The impact of the antiproton constraints is less visible in this case, as the relevant regions are already excluded by the new direct detection constraints. Hence, the top and bottom panels look almost identical. For both the propagation models, the likelihood of the best-fit is changed by $-2 \Delta \log \mathcal{L} \approx 2.5$, driven by the small excess seen in Higgs invisible decays. Thus, the coupling of the best-fit is also shifted to larger values ($\sim 0.02$). In the resonance region, the allowed parameter space is constrained from the left by the improved direct detection constraints, which also exclude a large region of the parameter space in the high-mass region.
The remaining parameter space will be further explored by future direct DM detection experiments~\cite{Aalbers:2022dzr}.

\section{Conclusions}

Observations of cosmic rays (CRs), in particular antiparticle fluxes, are among the most sensitive probes of dark matter (DM) annihilations, provided that the uncertainties related to injection and propagation are under control. While various codes exist to simulate CR propagation, the computational cost in combination with the complexity of the underlying propagation models make it very difficult to comprehensively explore indirect detection constraints for different DM models. In this work we have addressed this challenge for the case of the AMS-02 antiproton data by combining two key ingredients: neural networks capable of accurately predicting primary and secondary antiproton fluxes, and efficient marginalisation of propagation uncertainties in the calculation of AMS-02 likelihoods. The required software tools, \textsf{DarkRayNet.v2} and \textsf{pbarlike}, are publicly available and can be readily interfaced with the most recent release of the \textsf{GAMBIT} global fitting framework, which automates the calculation of cross sections and branching ratios from Lagrangian-level inputs~\cite{Bloor:2021gtp}.

Using importance sampling, the calculation of the marginalised likelihood for a specific DM hypothesis requires calculating the primary and secondary antiproton fluxes and the resulting AMS-02 likelihood for approximately $10^4$ different combinations of propagation and solar modulation parameters. In our approach, this calculation takes approximately $10\,\mathrm{s}$ on a single cpu, which is sufficiently fast to include antiproton constraints in global fits of DM models. As an illustration, we have considered the scalar singlet DM model, updating all relevant constraints to include the most recent experimental results. While in principle this model can give sizeable signals in indirect detection experiments, the relevant high-mass regions of parameter space are largely excluded by the most recent direct detection experiments. Given these constraints, as well as a slight excess in LHC measurements of the Higgs invisible branching ratio, the resonance region remains as the parameter region of primary interest for this model. 

Our work also highlights the need for flexible propagation models and an accurate treatment of experimental uncertainties. All our analyses have been carried out using two different propagation models, which introduce a break in the injection spectrum and in the diffusion coefficient, respectively. We show that, for the case of annihilation into $b\bar{b}$, the former model gives rise to a slight excess for DM masses around $100\,\mathrm{GeV}$ and tight constraints on TeV-scale DM, while the latter yields an excess for DM masses in the TeV range and very strong constraints on DM masses below $100\,\mathrm{GeV}$. Once correlations in AMS-02 data are taken into account, the local significance of any of these excesses is well below $3\sigma$.

At present there is hence no evidence for a DM signal in AMS-02 data and only upper bounds on the interactions of DM particles can be obtained. Nevertheless, our analysis clearly demonstrates the strong potential of using the antiproton flux to search for DM. The software tools that we release together with this work make it possible to fully exploit this potential and efficiently reinterpret experimental data for a wide range of DM models. These methods will become more and more important as our understanding of CR propagation continues to improve, flux measurements for heavier antiparticles are released (e.g.\ by the GAPS balloon mission~\cite{Aramaki:2015laa}), and a new generation of indirect detection experiments begin to take data~\cite{Battiston:2021org,Schael:2019lvx}.

\acknowledgments

We would like to thank Michael Kr\"{a}mer for discussions and Tom\'{a}s Gonzalo for help with the GAMBIT interface. SB acknowledges the support by the Doctoral School 'Karlsruhe School of Elementary and
Astroparticle Physics: Science and Technology'. FK is partially funded by the Deutsche
Forschungsgemeinschaft (DFG) through the Emmy Noether Grant No. KA 4662/1-1. 
MK is supported by the Swedish Research Council under contracts 2019-05135 and 2022-04283 and the European Research Council under grant 742104. SM acknowledges the European Union's Horizon Europe research and innovation programme for support under the Marie Sklodowska-Curie Action HE MSCA PF–2021,  grant agreement No.10106280, project \textit{VerSi}.
This project used computing resources from the Swedish National Infrastructure for Computing (SNIC) under project Nos. 2021/3-42, 2021/6-326 and 2021-1-24 partially funded by the Swedish Research Council through grant no. 2018-05973. The training of the neural networks was performed with computing resources granted by RWTH Aachen University under project `rwth0754'. 

\appendix

\section{Details on new software tools}

\subsection{\textsf{DarkRayNet.v2}}
\label{app:DRN}

The \textsf{DarkRayNet} is a neural emulation tool based on recurrent neural networks, specifically LSTMs \cite{LSTM}, that can quickly simulate antiprotons, protons, and Helium CR spectra at Earth, based on \textsf{Galprop} simulations. It was first published in Ref.~\cite{2107.12395} and is designed to predict measurable CR spectra for different parameters in seconds, allowing quick scans for indirect DM searches. The neural network architecture of the \textsf{DarkRayNet} is designed to handle the large set of relevant parameters for CR propagation and the DM model parameters individually, and convert the physical parameters into the resulting cosmic-ray spectra, using multiple densely connected layers and finally an LSTM layer. 
The network was trained using a large set of fluxes for each particle type and origin (primary/secondary/DM annihilation). The samples in the training data were chosen from \textsf{MultiNest} scans used to evaluate the compatible parameter space with recent AMS-02 data for the parameters describing CR transport (see also section~\ref{sec:Multinest}). The network accuracy for predictions within the trained parameter regions is always sufficiently high, such that a difference between the emulated spectra and the corresponding conventionally simulated spectra from \textsf{Galprop} will always be significantly below the uncertainty in the most recent AMS-02 data. As a result, no systematic will be introduced when using this emulator for likelihood evaluations. The simulation of cosmic-ray spectra with \textsf{DarkRayNet} is about two to three orders of magnitude faster than with the standard \textsf{Galprop} setup, depending on the number of different cosmic-ray species considered.

In this paper, we have updated the model for CR propagation (INJ.BRK) with respect to the setup in Ref.~\cite{2107.12395}, and included a new model (DIFF.BRK), as described in section~\ref{sec:models}. 
With respect to the original release, the new version \textsf{v2}\footnote{\href{https://github.com/kathrinnp/DarkRayNet}{https://github.com/kathrinnp/DarkRayNet}} of \textsf{DarkRayNet} has thus been extended with newly trained networks for the updated models.

\subsection{\textsf{pbarlike}}

\textsf{pbarlike}\footnote{\href{https://github.com/sowmiya-balan/pbarlike}{https://github.com/sowmiya-balan/pbarlike}} is an open-source python code developed for indirect DM searches with antiprotons. One of the main aims of \textsf{pbarlike} is state-of-the-art treatment of antiproton production cross section uncertainties and data correlations; the other is to interface the powerful CR emulator, \textsf{DarkRayNet} with the global fitting framework \textsf{GAMBIT}. In addition, \textsf{pbarlike} also computes solar modulation of the antiproton flux. \textsf{pbarlike} treats solar modulation using the \textit{Force Field Approximation}. It currently includes the 7-year AMS-02 antiproton dataset \cite{Aguilar:2021tos} and is easily extendable to include more datasets. For likelihood calculations, one can choose to either marginalize over all nuisance parameters from CR propagation and solar modulation, or to profile over only the solar modulation parameters, keeping the propagation parameters fixed. For marginalization, the \textsf{MultiNest} sample of all nuisance parameters (described in sec. \ref{sec:Multinest}) is made available; for profiling, custom CR propagation parameters need to be provided. 

For a chosen propagation model (options are those available within \textsf{DarkRayNet}, i.e.\ DIFF.BRK and INJ.BRK), and a given set of CR propagation and DM model parameters, \textsf{pbarlike} obtains predictions for antiproton flux at the local interstellar region from \textsf{DarkRayNet}. It then simulates solar modulation of this flux and finally compares it against the given dataset to calculate likelihoods. \textsf{pbarlike} can be easily installed and used as either a standalone for antiproton constraints or in conjunction with \textsf{GAMBIT} for global fits. The complete documentation can be found at \href{https://pbarlike.readthedocs.io}{https://pbarlike.readthedocs.io}.

\subsection{GAMBIT interface}

The interface between \textsf{pbarlike} and \textsf{GAMBIT} (also called the \emph{frontend}) makes it possible to automatically evaluate the AMS-02 antiproton likelihood for any DM model implemented in \textsf{GAMBIT}. For any such model, all relevant annihilation cross sections are provided by the \texttt{ProcessCatalog} object~\cite{GAMBITDarkMatterWorkgroup:2017fax}, such that the branching fractions for different final states and the total annihilation cross section can be readily calculated. For new models, the Process Catalogue may be automatically generated from Lagrangian-level input using \textsf{CalcHEP}~\cite{Belyaev:2012qa} via the \textsf{GAMBIT Universal Model Machine}~\cite{Bloor:2021gtp}.

\section{Comparison of correlated and uncorrelated likelihoods}
\label{app:uncorr}

\begin{figure}[t]
\includegraphics[width=\textwidth]{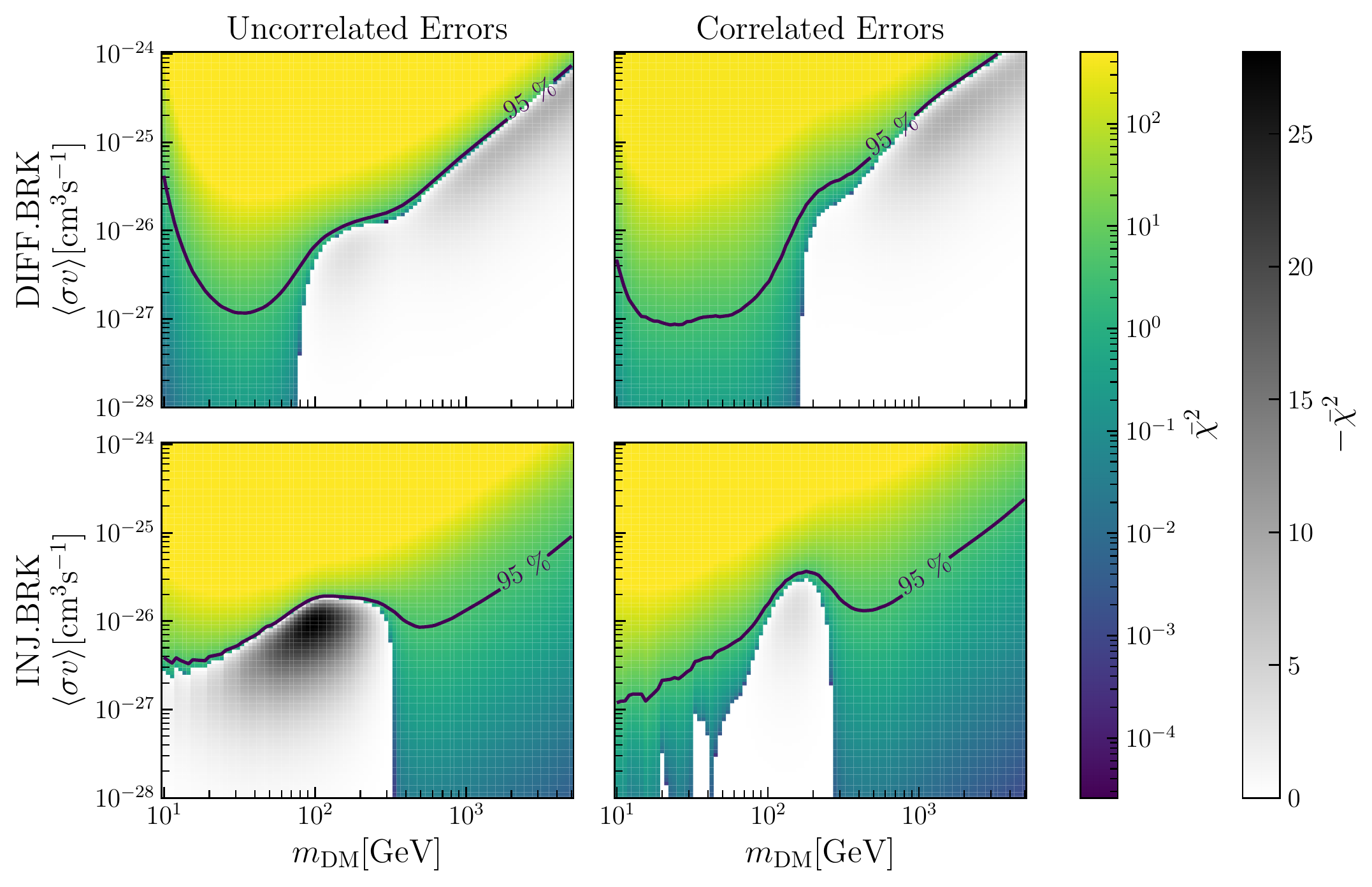}
		\caption{ 
            $\Bar{\chi}^2$ for DM annihilating into $b \Bar{b}$, obtained using the two CR propagation models - DIFF.BRK (\textit{top}), INJ.BRK (\textit{bottom}) - and using uncorrelated (left) and correlated (right) errors. 
        }
		\label{fig:bb bounds_all}
\end{figure}

Figure \ref{fig:bb bounds_all} compares the result from section~\ref{subsec: bb bounds} to the ones obtained when using uncorrelated errors. It can be observed that the inclusion of correlations leads to a decrease in local significance from $3.9 \sigma$ to $1.0 \sigma$ for the INJ.BRK model (assuming that $- \bar{\chi}^2_\text{min}$ follows a $\chi^2$ distribution with two degrees of freedom). The same effect is not observed in DIFF.BRK model where the local significance does not change by a lot when moving from uncorrelated ($2.3 \sigma$) to correlated ($2.2 \sigma$) errors, the reason being that errors are dominated by correlated systematic uncertainties only in the low rigidity range. 

\providecommand{\href}[2]{#2}\begingroup\raggedright\endgroup

\end{document}